\documentclass[twocolumn,showpacs,preprintnumbers,amsmath,amssymb]{revtex4}

\usepackage{graphicx}
\usepackage{dcolumn}
\usepackage{bm}

\begin{document}


\title{A lattice study of interaction mechanisms in a heavy-light
meson-meson system}

\author{Merritt S. Cook and H. Rudolf Fiebig}
\email{fiebig@fiu.edu}
\affiliation{Physics Department, FIU, 11200 SW 8th Street, Miami, Florida 33199, USA}

\date{\today}

\begin{abstract}
We study mass spectra of a meson-meson system involving two
light and two heavy quarks on an anisotropic lattice.
The heavy quarks are treated in the static approximation. The dependence of
the spectrum on the relative distance of the heavy quarks is extracted
from the lattice simulation using the maximum entropy method (MEM).
A correlation matrix of meson-meson operators
emphasizing quark and gluon exchange degrees of freedom is employed in an attempt
to learn about aspects of mechanisms of hadronic interaction.
\end{abstract}
\pacs{12.38.Gc, 12.40.Yx, 24.85.+p}

\maketitle

\section{\label{sec:intro}Introduction}
The term ``lattice hadron physics'' has been coined for
strong interaction physics based on first principles,
i.e. quantum chromodynamics (QCD).
In recent years hadron physics has emerged as a field in its own
right \cite{Capstick:2000dk}.
The desire to explain strong interaction phenomena in terms of the
underlying dynamics of quarks and gluons sets the field apart from
traditional nuclear physics, which emphasizes an
effective field theory point of view. 
Among the important issues faced by hadron physics are baryon and meson
spectroscopy, and structure, as well as the mechanism of their strong interaction.
Those issues have in common the need to deal with excited states.
In lattice QCD, which affords the most direct access to hadron physics at nuclear
energy scales, excited states spectroscopy is just now moving
into reach due to the use of anisotropic lattices, advanced analysis
techniques, and powerful computing facilities.

This work is concerned with learning about hadronic interaction mechanisms.
We believe that much of the physics of hadronic interaction can be understood
by investigating heavy-light systems. In those the relative distance of hadrons
is a well defined quantity. The lattice `data' can be interpreted in terms of
intuitive pictures, like potentials.
Insight into mechanisms of the strong interaction flows from looking at
excitations due to hadron-hadron operators, say $\Phi(t)$, at various
relative distances $r$. Different choices for the structure of those operators
in terms of their composition from quark and gluon fields may potentially point
at interesting physics of the system, such as the importance of
quark versus gluon exchange degrees of freedom as a function of $r$.
To extract information of this kind the computation of matrix elements
$\langle n|\Phi(t_0)|0\rangle$ between the vacuum $|0\rangle$ and
ground and excited states $|n\rangle$, $n>0$, will be useful.

From the vantage point of a numerical lattice simulation this can be a notoriously
difficult problem. Hadron-hadron operators are prone to produce very noisy
correlation functions. Extracting spectral information in the standard
fashion, i.e. trying to identify a plateau in an effective mass function,
may not be practical.

An alternative analysis method that is just being `discovered' by lattice
practitioners is application of the maximum entropy method (MEM) or an otherwise
constrained form of Bayesian inference
\cite{Nakahara:1999bm,Lepage:2001ym,Yamazaki:2001er}.
From the Bayesian perspective the parameters of a model for
the Euclidean time correlation function are viewed as random variables
drawn from a certain probability distribution function. The latter, known
as the posterior probability ${\cal P}[\rho\leftarrow C]$, is the conditional
probability for a certain parameter set $\rho$ given a data set $C$.
In our case the data set $C$ is the lattice-measured time correlation function
$C(t,t_0)$ and the parameter set $\rho$ is the spectral density function
in the model
\begin{equation}
F(\rho|t,t_0)=\int d\omega\,\rho(\omega)\,e^{-\omega(t-t_0)}\,.
\label{Fc}\end{equation}
Discretization understood, in this approach the number of parameters is
allowed to exceed
the number of data points without causing conceptual problems.
In the MEM the posterior probability is constructed from the $\chi^2$-distance between
the lattice data $C$ and the model $F$, and the information
content of $\rho$ measured by the entropy $S=-\int d\omega\rho(\omega)\ln\rho(\omega)$.
Usually, the result inferred from the Bayesian approach is the {\em most likely}
parameter set $\rho$.

Analytically, the spectral density function is a sum of discrete $\delta$-peaks
\begin{equation}
\rho(\omega)=
\sum_{n\neq 0}\delta(\omega-\omega_n)\,
|\langle n|\Phi(t_0)|0\rangle|^2\,.
\label{rhodelta}\end{equation}
From a numerical viewpoint computational constraints on $C$ render the peak widths
finite. Physical quantities are contained in each peak as low $\omega$ moments.
Among those are the peak volume $|\langle n|\hat{\Phi}(t_0)|0\rangle|^2$
and the peak energy $E_n$, i.e. the mean value of $\omega$.

The Bayesian spectral analysis of the lattice data is an interesting problem
in itself. It leads to
the discussion of a number of computational strategies in a general context.
In order to keep this presentation focused, we refer the reader to
a separate paper \cite{Fiebig:2002sp} where selected aspects of Bayesian spectral analysis
are discussed, using the same lattice simulation data as a testing ground.

Starting a little more than a decade ago lattice work on hadronic interaction has
followed mostly two tracks; investigation of heavy-light systems for varying
hadron relative distance \cite{Richards:1990xf},
and the computation of scattering phase shifts from energy spectra of two hadrons in
a finite box \cite{Luscher:1991ux}.
We will not review these subjects here, but rather
mention a few leads to facilitate following the
literature.\footnote{For reviews see \protect\cite{Opus5:2002,Green:2002}}
Extraction of scattering phase shifts has been successful in terms of structureless
particles \cite{Goe94}. On the other hand, scattering of composite hadrons within
L\"uscher's framework is considerably more difficult. Scattering phase shifts
have been obtained for the S-wave interaction of the $\pi\pi$ system
in the $I=2$ channel \cite{Fiebig:1999hs}, and by the CP-PACS
collaboration \cite{Aoki:2001hc}.
Considerable computing power was brought to bear by the JLQCD collaboration
to extracting scattering lengths for the $\pi\pi$, $\pi$N, and NN systems
\cite{Fukugita:1995ve}. While lattice volume limitations hinder a realistic
treatment of the NN system,
the $\pi$ scattering lengths of Ref.~\cite{Fukugita:1995ve} stand out as the only
quantitative results for hadronic interaction from the lattice to date.
Systems with heavy, even static, quarks will not yield quantitative results,
but are useful to gain a deeper understanding of the mechanisms that lead
to the phenomenology of nuclear forces. Green and coworkers have been
studying energies of static four-quark systems in
various geometric configurations, most recently with two static and two light
quarks \cite{Green:1999mf}. Their results shed light on the importance of many-body
versus two-body forces in hadrons. Other studies of heavy-light systems
\cite{Mihaly:1997ue} focus on adiabatic potentials, i.e. the ground state energies as
functions of the relative hadron-hadron separation.
Along these lines, most of the hadronic interaction work by members of the UKQCD
collaboration can be traced from \cite{Michael:1999nq}. 
There, heavy-light meson-meson systems are classified according to
isospin and spin configurations of the light quarks (${\cal B}$ mesons).
Depending on the channel, interaction potentials at around $\approx 0.5{\rm fm}$
come out attractive or repulsive, with magnitudes rarely exceeding $\approx 50{\rm MeV}$.
Though this is typical for nuclear physics, from the point of view of a lattice
simulation these are very small energy differences to be measured.
This is a generic problem for hadronic interaction physics on the lattice.

In the present work we report on a study of a system of two heavy-light
mesons based on two operators. The first one, $\Phi_1$, is the standard product
of two local pseudoscalar heavy-light meson
fields at distance $r$. The other one, $\Phi_2$, is similar, but nonlocal, probing color
rearrangement. In some sense $\Phi_1$ and $\Phi_2$ test the
importance of quark and gluon exchange degrees of freedom for the interaction.
They enter into a $2\times 2$ time correlation matrix.
Our goal here is to learn about the interaction mechanisms represented by those
operators as a function of the relative meson-meson distance.
The current work goes beyond previous studies of heavy-light meson-meson
systems mainly in that (i) nonlocal operators are considered and (ii) operator mixing is
built into time correlation functions, allowing the system to `dynamically pick' the
dominant excitation mechanism as the meson-meson separation $r$ varies.
Moreover, due to the spectral analysis used in this work we are able to (iii)
extract the actual excitation strengths of the ground and excited two-meson states.

Preliminary results reported in \cite{Fiebig:2001mr,Fiebig:2001nn} were
based on an analysis of the diagonal correlator elements, with no mixing.
There, averaging over annealing start configurations
had not been done, the spectral density functions came from single annealing runs.
As it turns out this is a source of systematic error that can not be tolerated in
the light of the smallness of the extracted energy shifts.

\section{\label{sec:action}Lattice action}

The meson-meson operators employed in this simulation lead to somewhat massive states.
The resulting steep drop of time correlation functions,
particularly for the excited states, makes it very difficult
to analyze the lattice signal.
To deal with this situation we use an anisotropic lattice action.
If the aspect ratio $\xi=a_s/a_t$
of the lattice constants in space and time directions, respectively, is made
larger than one the number of usable time correlation function data is increased
before the signal `vanishes' into noise.
Anisotropic lattices have been essential for computing the glue ball
mass spectrum \cite{Morningstar:1999rf}. The current simulation of hadronic
interaction has in common the need for extracting excited states.

The gauge field part of the lattice action has the form
\begin{subequations}
\begin{eqnarray}
S_G[U]&=&\beta\sum_{\ell} c_\ell \Omega_\ell\quad{\rm with}\quad\label{SG}\\
\Omega_\ell&=&\sum_{{\cal C}\in{\cal S}_\ell}
\frac13{\rm Re}{\rm Tr}[\openone-U({\cal C})]\label{SO}\,.
\end{eqnarray}
\end{subequations}
Here $\beta=6/g^2$ in terms of the gauge coupling $g$,
$\ell$ labels sets ${\cal S}_\ell$ of closed lattice contours ${\cal C}$
and $U({\cal C})$ denotes the path ordered product of gauge field link variables
along ${\cal C}$. We adopt a tree-level tadpole improvement scheme with
four classes of loops: all oriented spatial elementary plaquettes $\ell=ss$,
temporal elementary plaquettes $\ell=st$, spatial planar rectangles
$\ell=sss$, and short temporal planar rectangles with two spatial and one
temporal link $\ell=sst$. Specifically
\begin{eqnarray}
S_G[U]&=&\beta\left\{ \frac{5}{3\xi u_s^4}\Omega_{ss}
          +\frac{4\xi}{3 u_s^2 u_t^2}\Omega_{st} \right. \nonumber\\ 
 & & \left. -\frac{1}{12\xi u_s^6}\Omega_{sss}
          -\frac{\xi}{12 u_s^4 u_t^2}\Omega_{sst} \right\}\,, \label{SG4}
\end{eqnarray}
where $u_s$ and $u_t$ are spatial and temporal link renormalization
factors \cite{Lepage:1993xa}, and
\begin{eqnarray}
\Omega_{ss}&=&\sum_x\sum_{1\leq\mu<\nu\leq3}\frac13{\rm Re}{\rm Tr}[\openone-
U_\mu(x)\times\nonumber\\
&&U_\nu(x+\hat{\mu})U^\dagger_\mu(x+\hat{\nu})U^\dagger_\nu(x)] \label{Uss}\\
\Omega_{st}&=&\sum_x\sum_{1\leq\mu\leq3}\frac13{\rm Re}{\rm Tr}[\openone-
U_\mu(x)\times\nonumber\\
&&U_4(x+\hat{\mu})U^\dagger_\mu(x+\hat{4})U^\dagger_4(x)] \label{Ust}\\
\Omega_{sss}&=&\sum_x\sum_{1\leq\mu\neq\nu\leq3}\frac13{\rm Re}{\rm Tr}[\openone
-U_\mu(x)U_\mu(x+\hat{\mu})\times\nonumber\\
&& U_\nu(x+2\hat{\mu})U^\dagger_\mu(x+\hat{\nu}+\hat{\mu})
U^\dagger_\mu(x+\hat{\nu})U^\dagger_\nu(x)] \label{Usss}\\
\Omega_{sst}&=&\sum_x\sum_{1\leq\mu\leq3}\frac13{\rm Re}{\rm Tr}[\openone
-U_\mu(x)U_\mu(x+\hat{\mu})\times\nonumber\\
&& U_4(x+2\hat{\mu})U^\dagger_\mu(x+\hat{4}+\hat{\mu})
U^\dagger_\mu(x+\hat{4})U^\dagger_4(x)]\,.\phantom{X} \label{Usst}\,.
\end{eqnarray}
This action is the same as in \cite{Morningstar:1999rf}, the chosen
parameters are $\beta=2.4$ and $\xi=3$.
The spatial link renormalization is
computed self consistently from the average spatial plaquette, 
\begin{equation}
u_s=\left(1-\langle\Omega_{ss}\rangle/3L^3\right)^{1/4}\,,
\label{us}\end{equation}
while the temporal renormalization factor is set to one, $u_t=1$.
According to \cite{Morningstar:1999rf} this results in a spatial lattice
constant of about $a_s\approx 0.25{\rm fm}$, or $a_s^{-1}\approx 800{\rm MeV}$.

The fermion action is
\begin{equation}
S_F[U,\psi,\bar{\psi}]=\sum_{x,y} \bar{\psi}_{fA\mu}(x)
Q_{fA\mu,gB\nu}(x,y) \psi_{gB\nu}(y)\,,
\label{SF} \end{equation}
with $f,g$=flavor, $A,B$=color, and $\mu,\nu$=Dirac indices.
The fermion matrix $Q$ is assumed flavor diagonal
\begin{equation}
Q_{fA\mu,gB\nu}(x,y)=\delta_{fg} Q^{(f)}_{A\mu,B\nu}(x,y)\,,
\label{Qf} \end{equation}
and has a Wilson and a clover term \cite{Wilson:1974sk,Sheikholeslami:1985ij}
\begin{equation}
Q^{(f)}=\openone-\kappa^{(f)}(M-c_{\rm SW}K)\,.
\label{QMK} \end{equation}
In detail
\begin{eqnarray}
\lefteqn{M(x,y)=}\label{QM}\\
&&\frac{1}{u_s}\sum_{\mu=1}^3\left[
 (r_s-\gamma_\mu)U_\mu(x)\delta_{x+\hat{\mu},y}
+(r_s+\gamma_\mu)U^\dagger_\mu(y)\delta_{x,y+\hat{\mu}}\right]\nonumber\\
&& +\frac{\xi}{u_t}[(r_t-\gamma_4)U_4(x)\delta_{x+\hat{4},y}
   +(r_t+\gamma_4)U^\dagger_4(y)\delta_{x,y+\hat{4}}]\,.\nonumber
\end{eqnarray}
The critical hopping parameter in the anisotropic case is
$\kappa_c=(2\xi r_t+6 r_s)^{-1}$. The Wilson parameters are chosen as
$r_s=r_t=1$. Because the lattice is coarse in the space directions and,
more importantly, because we wish to avoid problems with ghosts (unphysical
branches in the lattice-quark dispersion relation \cite{Alford:1997nx})
we use clover improvement \cite{Sheikholeslami:1985ij} only in the spatial planes,
\begin{equation}
K(x,y)=\delta_{x,y}\frac{1}{u_s^4}\sum_{1\leq\mu<\nu\leq3}\sigma_{\mu\nu}
\frac{1}{2i}\left(P_{\mu\nu}(x)-P^\dagger_{\mu\nu}(x)\right)\,.
\label{QK}\end{equation}
Here $\sigma_{\mu\nu}=\frac{i}{2}[\gamma_\mu,\gamma_\nu]$ and
$P_{\mu\nu}(x)=\frac14\sum_{k=1}^4P_{k,\,\mu\nu}(x)$ is made from four
transport operators along oriented 4-link paths in the $\mu$--$\nu$ plane,
starting and terminating at $x$, for example
$P_{1,\,\mu\nu}(x)=U_\mu(x)U_\nu(x+\hat{\mu})U^\dagger_\mu(x+\hat{\nu})U^\dagger_\nu(x)$,
see \cite{Sheikholeslami:1985ij}.

The strength coefficient is $c_{\rm sw}=1$.
At $\kappa^{(f)}=0.0679$ we have $m_\pi/m_\rho\approx 0.75$ on
a $L^3\times T=10^3\times 30$ lattice.
Using $m_{\rm eff}a_t=0.4676$ from \cite{Fiebig:2002sp} a crude estimate for the
quark mass puts it within 15\% above the strange mass scale.
We have used a hybrid molecular dynamics algorithm (HMC) \cite{Duane:1986iw} to
generate $N_U=708$ quenched gauge configurations.

\section{\label{sec:operators}Operators}

At this time we wish to address the physical
mechanisms responsible for the features of hadronic interaction rather than
making quantitatively precise predictions.
Toward this end we employ, for a two-meson system, a set of operators
meant to excite different QCD degrees of freedom.
For two heavy-light pseudoscalar mesons, for example, the operator
\begin{equation}
\Phi_1(t)=\sum_{\vec{x},\vec{y}}
\delta_{\vec{r}\,,\vec{x}-\vec{y}}\,
\overline{Q}_A(\vec{x}t) \gamma_5 q_A(\vec{x}t)\,
\overline{Q}_B(\vec{y}t) \gamma_5 q_B(\vec{y}t)\,
\label{Phi1}\end{equation}
is built from separable products of two color singlets,
describing two mesons at relative distance $\vec{r}$.
In (\ref{Phi1}) $Q$ and $q$ are the heavy and light quark fields, respectively, 
and $A,B$ are color indices.
Color contractions are done between quark fields which spatially coincide at
$\vec{x}$ and $\vec{y}$, respectively.
So far, only local operators of this kind have been employed for hadronic
interaction studies \cite{Fiebig:1999hs,Michael:1999nq,Mihaly:1997ue,Richards:1990xf}.
At small values of $r$ color coupling schemes involving quarks in different
mesons may become dynamically possible. An operator testing excitations of that nature is
\begin{eqnarray}
\Phi_2(t)&=&\sum_{\vec{x},\vec{y}}
\delta_{\vec{r}\,,\vec{x}-\vec{y}}\,
U_{P;AA^\prime}(\vec{x}t,\vec{y}t)\,
U^\dagger_{P^\prime;B^\prime B}(\vec{x}t,\vec{y}t)\nonumber\\
&&\overline{Q}_A(\vec{x}t) \gamma_5 q_B(\vec{x}t)\,
\overline{Q}_{B^\prime}(\vec{y}t) \gamma_5 q_{A^\prime}(\vec{y}t)\,.
\label{Phi2}\end{eqnarray}
It involves link products $U_P(x,y)$ along spatial paths $P$ within a fixed time slice.
The operator $\Phi_2(t)$ interpolates fields that are still products of two color
singlets, but those are now formed from quarks at different locations $\vec{x}$
and $\vec{y}$.

In order to simplify the notation we will tacitly assume dependence of
all subsequent quantities on the relative distance $\vec{r}$, but suppress
$\vec{r}$ in most of the expressions below.

We are thus lead to computing the elements
\begin{equation}
C_{ij}(t,t_0)=
\langle\hat{\Phi}_{i}^\dagger(t)\hat{\Phi}_{j}(t_0)\rangle,\quad i,j=1,2\,,
\label{C22}\end{equation}
of $2\times 2$ time correlation matrices, one for each $\vec{r}$, 
where $\hat{\Phi}=\Phi-\langle \Phi\rangle$, means vacuum-subtracted operators.
The correlation matrix $C(t,t_0)$ built from the operators (\ref{Phi1}) and
(\ref{Phi2}) can be worked out by way of Wick's theorem, symbolically
\begin{widetext}
\begin{eqnarray}
\bar{Q}q\bar{Q}q\,\bar{q}Q\bar{q}Q &=&
 \stackrel{4}{\rule{0mm}{4mm}\bar{Q}}\stackrel{3}{\rule{0mm}{4mm}q}
\stackrel{2}{\rule{0mm}{4mm}\bar{Q}}\stackrel{1}{\rule{0mm}{4mm}q}\,
\stackrel{1}{\rule{0mm}{4mm}\bar{q}}\stackrel{2}{\rule{0mm}{4mm}Q}
\stackrel{3}{\rule{0mm}{4mm}\bar{q}}\stackrel{4}{\rule{0mm}{4mm}Q}
+\stackrel{4}{\rule{0mm}{4mm}\bar{Q}}\stackrel{3}{\rule{0mm}{4mm}q}
\stackrel{2}{\rule{0mm}{4mm}\bar{Q}}\stackrel{1}{\rule{0mm}{4mm}q}\,
\stackrel{1}{\rule{0mm}{4mm}\bar{q}}\stackrel{4}{\rule{0mm}{4mm}Q}
\stackrel{3}{\rule{0mm}{4mm}\bar{q}}\stackrel{2}{\rule{0mm}{4mm}Q}
+\stackrel{4}{\rule{0mm}{4mm}\bar{Q}}\stackrel{3}{\rule{0mm}{4mm}q}
\stackrel{2}{\rule{0mm}{4mm}\bar{Q}}\stackrel{1}{\rule{0mm}{4mm}q}\,
\stackrel{3}{\rule{0mm}{4mm}\bar{q}}\stackrel{2}{\rule{0mm}{4mm}Q}
\stackrel{1}{\rule{0mm}{4mm}\bar{q}}\stackrel{4}{\rule{0mm}{4mm}Q}
+\stackrel{4}{\rule{0mm}{4mm}\bar{Q}}\stackrel{3}{\rule{0mm}{4mm}q}
\stackrel{2}{\rule{0mm}{4mm}\bar{Q}}\stackrel{1}{\rule{0mm}{4mm}q}\,
\stackrel{3}{\rule{0mm}{4mm}\bar{q}}\stackrel{4}{\rule{0mm}{4mm}Q}
\stackrel{1}{\rule{0mm}{4mm}\bar{q}}\stackrel{2}{\rule{0mm}{4mm}Q}
\,,
\label{Qq}\end{eqnarray}
where pairs $nn$ of numbers $n=1\ldots 4$ denote contractions. 
This leads to the following expression
\begin{eqnarray}
\lefteqn{ C_{ij}(t,t_0)=2\delta^{(+)}_{\vec{r},\vec{r}^{\,\prime}}
\langle \sum_{\vec{x},\vec{y}}
\delta^{(+)}_{\vec{r}\,,\vec{x}-\vec{y}}\,
{\cal U}_{i;AA^\prime,BB^\prime}(t,\vec{x}\vec{y})\,
{\cal U}_{j;DD^\prime,CC^{\prime}}(t_0,\vec{x}\vec{y}) } \nonumber\\
&&H^\ast_{B^\prime\nu^\prime,D^\prime\lambda^\prime}(\vec{y}t,\vec{y}t_0)
H^\ast_{A\nu,C\lambda}(\vec{x}t,\vec{x}t_0)
[G_{A^\prime\nu^\prime,C^\prime\lambda^\prime}(\vec{y}t,\vec{y}t_0)
G_{B\nu,D\lambda}(\vec{x}t,\vec{x}t_0)
-G_{A^\prime\nu^\prime,D\lambda}(\vec{y}t,\vec{x}t_0)
G_{B\nu,C^\prime\lambda^\prime}(\vec{x}t,\vec{y}t_0)]\rangle\,.\phantom{XX}
\label{Cij}\end{eqnarray}
\end{widetext}
In (\ref{Cij}) the symmetrized Kronecker symbol
\begin{equation}
\delta^{(+)}_{\vec{r},\vec{r}^{\,\prime}}=
\delta_{+\vec{r},\vec{r}^{\,\prime}}+\delta_{-\vec{r},\vec{r}^{\,\prime}}
\label{deltap}\end{equation}
is related to $O(3,{\mathbb Z})$ symmetry, projecting heavy-quark distances to
absolute lengths $r=|\vec{x}-\vec{y}|$.
The heavy anti-quark propagator is employed in the static approximation,
i.e. the leading term of the hopping parameter expansion\footnote{The
factor $(2\kappa)^{t-t_0}$ only produces an irrelevant mass shift. It has
been dropped from all subsequent correlator functions.}
\begin{equation}
H^\ast_{A\nu,C\lambda}(\vec{x}t,\vec{y}t_0)=
\delta_{\vec{x},\vec{y}}\,
(2\kappa)^{t-t_0}\frac12(\openone+\gamma_4)_{\lambda\nu}
U_{CA}(\vec{x}t_0,\vec{x}t)\,,
\label{Gheavy}\end{equation}
where $U(\vec{x}t_0,\vec{x}t)$ is the path-ordered link variable product along a straight
line from $\vec{x}t_0$ to $\vec{x}t$.
Finally, depending on whether $\Phi_1$ or $\Phi_2$ is involved in the correlator matrix
element, the contour operators
\begin{eqnarray}
{\cal U}_{1;AA^\prime,BB^\prime}(t,\vec{x}\vec{y})&=&
\delta_{AB}\delta_{A^\prime B^\prime} \label{U1cal}\\
{\cal U}_{2;AA^\prime,BB^\prime}(t,\vec{x}\vec{y})&=&
U_{P;AA^\prime}(\vec{x}t,\vec{y}t)
U^\ast_{P^\prime;BB^\prime}(\vec{x}t,\vec{y}t)\nonumber\\
&&\hspace{-2ex}+U_{P^\prime;AA^\prime}(\vec{x}t,\vec{y}t)
U^\ast_{P;BB^\prime}(\vec{x}t,\vec{y}t)\phantom{XX} \label{U2cal}
\end{eqnarray}
are trivial, or involve path-ordered link variable products along purely spatial
paths $P,P^\prime$ from $\vec{x}t$ to $\vec{y}t$.
On the present lattice we consider straight on-axis paths of lengths $r=1,2,3,4$.

\begin{figure}
\includegraphics[angle=0,width=72mm]{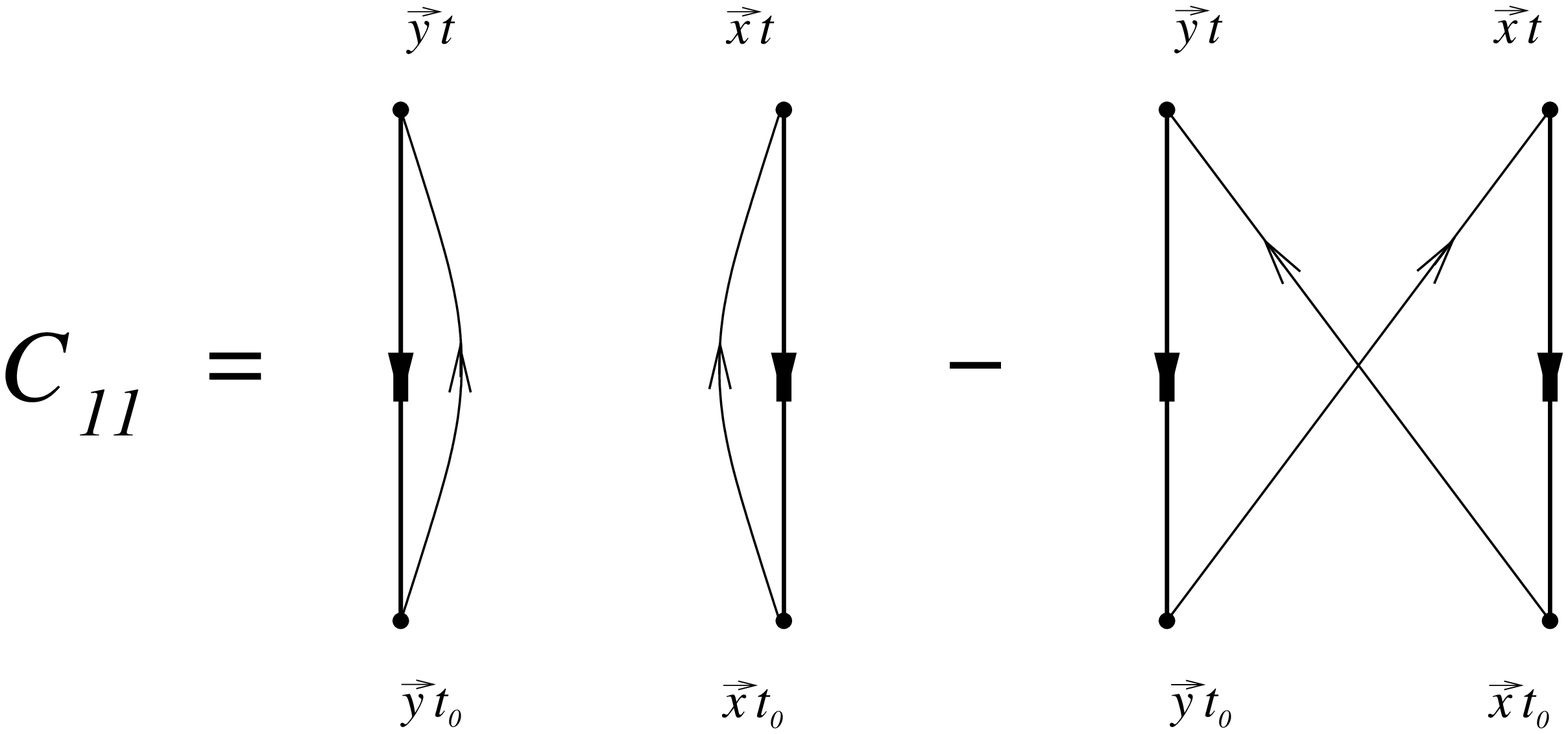}\vspace{2ex}\\
\includegraphics[angle=0,width=72mm]{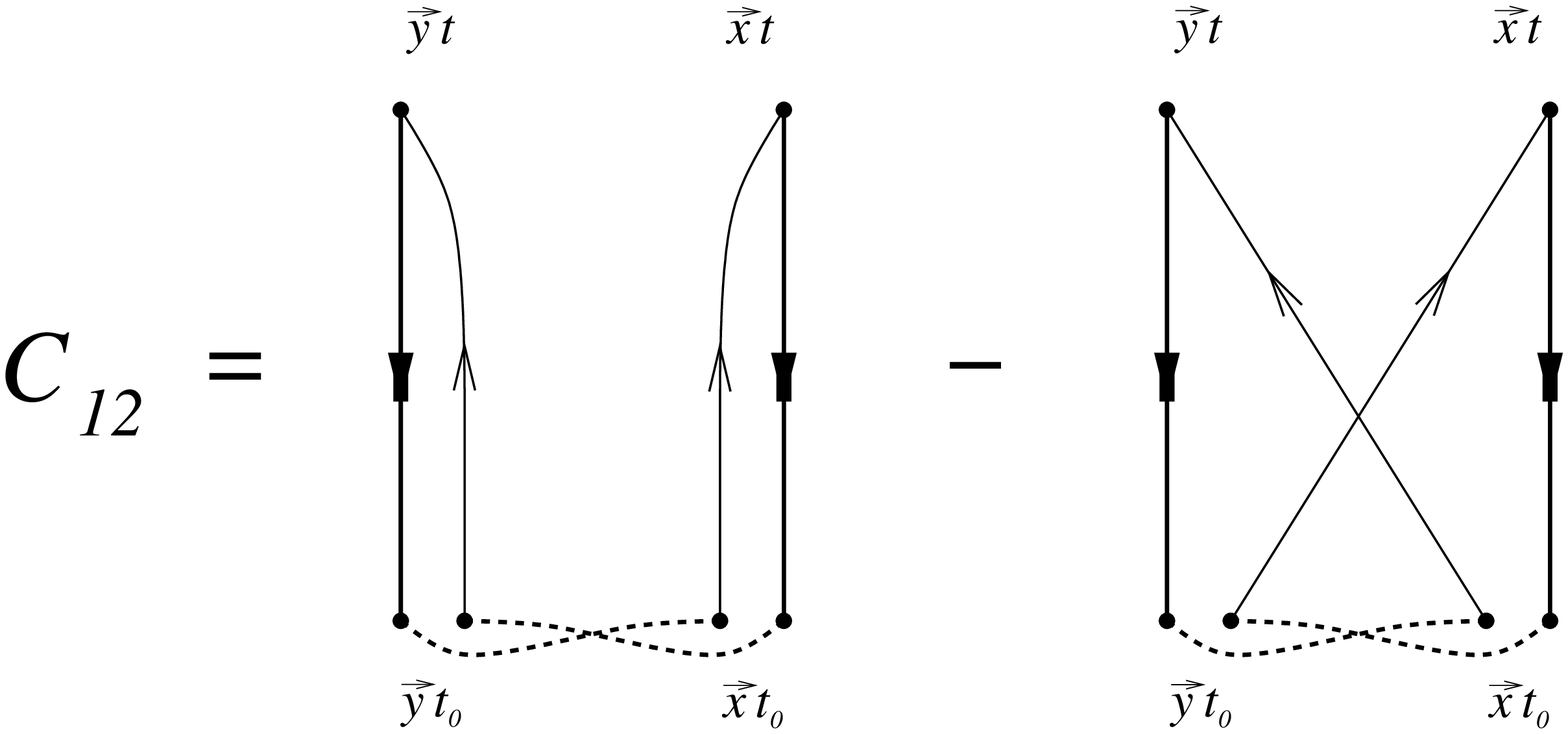}\vspace{2ex}\\
\includegraphics[angle=0,width=72mm]{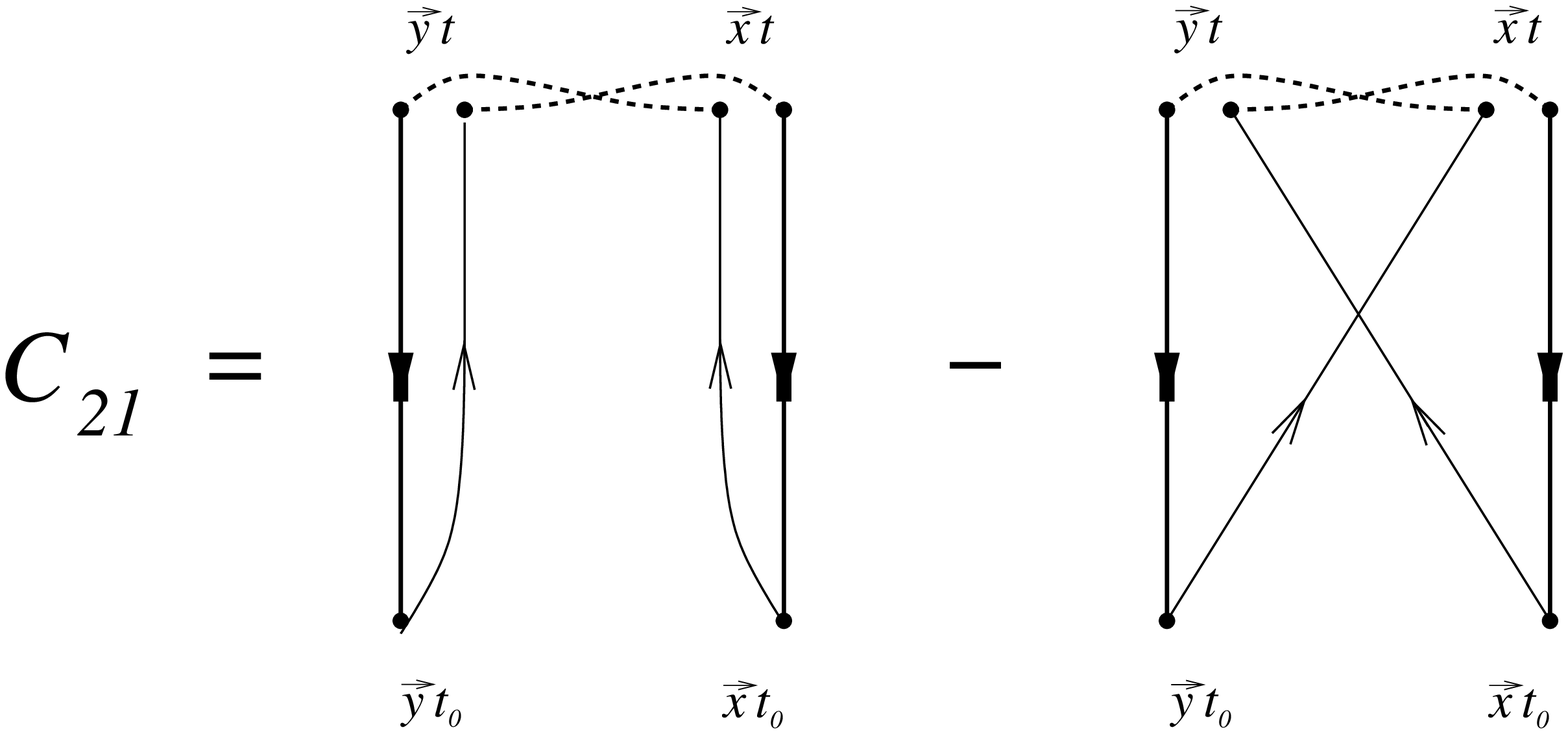}\vspace{2ex}\\
\includegraphics[angle=0,width=72mm]{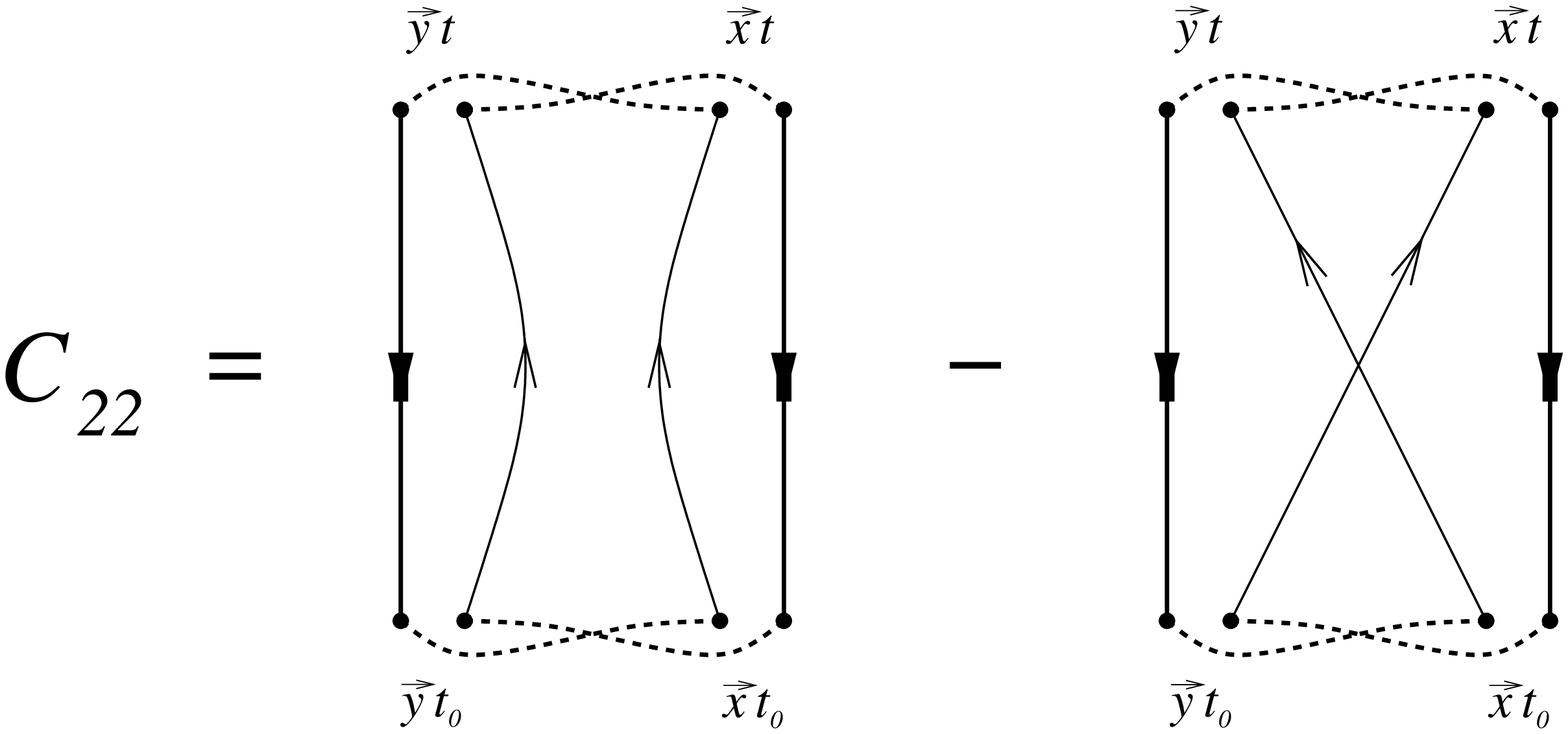}\vspace{2ex}
\caption{\label{fig1}Diagrammatic representation of the elements of the $2\times 2$
correlator matrix (\protect\ref{Cij}). Thick and thin solid lines mean heavy and light
quark propagators, respectively. Gluon propagation, i.e. products of link variables,
are shown as dashed lines. Pairs of dots refer to the same space-time point.}
\end{figure}

Diagrammatic likenesses of the correlator matrix elements (\ref{Cij}) are
shown in Fig.~\ref{fig1}. The simplest one, the graph of $C_{11}$, is the
standard quark exchange diagram that is usually
considered \cite{Fiebig:1999hs,Michael:1999nq,Mihaly:1997ue,Richards:1990xf},
while the rest involve explicit gluon degrees of freedom.

In (\ref{Cij}) light-quark propagator elements, e.g.
$G_{A^\prime\nu^\prime,D\lambda}(\vec{y}t,\vec{x}t_0)$,
between arbitrary lattice sites $\vec{x}t,\vec{y}t_0$ emerge.
The source time slice $t_0$ may be kept fixed, however, all-to-all spatial
propagator elements are needed.
More precisely, after working out one of the space-site sums in (\ref{Cij})
the number of propagator columns needed may be minimized by using translational
invariance of the gauge field average $\langle\ldots\rangle$. Then, the
number of propagator columns becomes equal to the number of relative distances
$\vec{r}$. On the other hand, employing all-to-all propagator elements has
the advantage of greater flexibility with regard to varied choices for $\vec{r}$,
like off-axis distances (future studies), and improved statistics due to
space-site averaging in (\ref{Cij}).
Because the gauge link contour operators (\ref{Gheavy}) and (\ref{U2cal}) tend
to make the correlation matrix (\ref{Cij}) quite noisy,
we follow the latter strategy. 

Random-source estimation is a proven technique for generating all-to-all propagators
\cite{Scalettar:1986uy,Duane:1986iw,Gottlieb:1987eg,Fiebig:1999hs,Michael:1998sg}.
Consider the following linear equation
\begin{eqnarray}
\sum_{\vec{y}y_4}\sum_{B\nu}
Q_{A\mu,B\nu}(\vec{x}x_4,\vec{y}y_4)
X^{(A^S\mu^Srx^S_4)}_{B\nu}(\vec{y}y_4)&=&\nonumber\\
&&\hspace{-31ex}\delta_{AA^S}\delta_{\mu\mu^S}
R^{(A^S\mu^Srx^S_4)}(\vec{x})\delta_{x_4x^S_4}\,.
\label{QXR}\end{eqnarray}
On the right-hand side $R^{(A^S\mu^Srx^S_4)}$ denotes a vector of length $L^3$
of complex random deviates.
Indices superscripted with ${}^S$ characterize the source
with respect to color, Dirac, and time slice. 
We choose sources that are nonzero on only one fixed time slice $x^S_4=t_0$,
whereas for each of the 12 color-Dirac combinations $A^S=1\ldots 3$,
$\mu^S=1\ldots 4$, in turn, $r=1\ldots N_R$ different random vectors 
are generated. In the simulation complex ${\mathbb Z}_2$ distributed random deviates
were used \cite{Dong:1994pk}. With $N_R=8$ this results in $96$ statistically
independent sources per gauge configuration. Writing
$\sum_{<r>}$ for the ensemble average, of which
$\frac{1}{N_R}\sum_{r=1}^{N_R}$ is a truncation, we have
\begin{equation}
\sum_{<r>}
R^{(A^S\mu^S rx^S_4)}(\vec{x})
R^{(A^S\mu^S rx^S_4)\ast}(\vec{y})
=\delta_{\vec{x}\vec{y}}
\label{RRdelta}\end{equation}
for all $A^S,\mu^S,x^S_4$ with appropriately chosen normalization.
Application of (\ref{RRdelta}) to (\ref{QXR}) yields
\begin{equation}
G_{B\nu,A\mu}(\vec{y}t,\vec{x}t_0)=
\sum_{<r>}
X^{(A\mu r t_0)}_{B\nu}(\vec{y}t)
R^{(A\mu rt_0)\ast}(\vec{x})
\label{GXR}\end{equation}
as a stochastic estimator for spatial all-to-all propagator matrix elements.

We have used the BiCGStab algorithm \cite{Frommer:1994vn,Vor92} for
solving (\ref{QXR}).

Operator smearing \cite{Alexandrou:1994ti} is a technique for enhancing the
amplitude of the ground state in a correlation function.
We smear the light quark fields only, defining iteratively
\begin{subequations}
\begin{eqnarray}
q^{\{ 0\}}_{A}(\vec{x}t)&=&q_{A}(x) \label{Kqa}\\
q^{\{ m\}}_{A}(\vec{x}t)&=&\sum_{B} \sum_{\vec{y}}
{\cal K}_{AB}(\vec{x},\vec{y}\,)\;q^{\{ m-1\}}_{B}(\vec{y}t) \,,
\label{Kqib}\end{eqnarray}
\end{subequations}
with $m\in {\mathbb N}$, and the matrix
\begin{eqnarray}
{\cal K}_{AB}(\vec{x},\vec{y}\,)&=&\delta_{AB}\,\delta_{\vec{x},\vec{y}}
+\alpha \sum_{\mu=1}^{3}
\left[\rule{0mm}{2.5ex}U_{\mu,AB}(\vec{x}\,t)\delta_{\vec{x},\vec{y}-\hat{\mu}}
\right.\nonumber\\
&&\left.+U^\dagger_{\mu,AB}(\vec{y}\,t)\delta_{\vec{x},\vec{y}+\hat{\mu}}
\right]\,.
\label{smear2}\end{eqnarray}
The real number $\alpha$ and the maximum value $M$ for iterations
$m=0\ldots M$ are parameters.  We have also used APE type \cite{Alb87a} fuzzy
link variables $U\in SU(3)$ in (\ref{smear2}) with the same parameter set.
The above iterative prescription
translates directly to the random source and solution vectors $R$ and $X$,
i.e. replacing $R\rightarrow R^{\{ M\}}$ and  $X\rightarrow X^{\{ M\}}$
in (\ref{GXR}) yields the propagator $G^{\{ M\}}$ for smeared fermion fields.
For more technical details see \cite{Fiebig:1999hs}.

Smearing was employed as a matter of course. In fact the operator (\ref{Phi2})
is designed with excited states of the two-meson system in mind.
Thus light-quark field smearing and link variable fuzzing may be of limited
value for the task at hand.
We have thus chosen somewhat conservative parameter values $\alpha=1.4$ and $M=4$ for
the current simulation.

\section{\label{sec:analysis}Spectral density analysis}

We now turn to the problem of extracting spectral information from the correlation
matrix (\ref{C22},\ref{Cij}). It has the decomposition
\begin{eqnarray}
C_{ij}(t,t_0)&=&\sum_{n\neq 0}
\phi_{ni} \phi_{nj}^\ast e^{-E_n(t-t_0)} \label{Cdcmp} \\
\phi_{ni}&=&\langle 0|\Phi_i^\dagger(t_0)|n\rangle\,, \label{Cphin}
\end{eqnarray}
where $|n\rangle$ denotes a complete set of states with energies $E_n$,
some of them may be negative in a lattice simulation.
The sum in (\ref{Cdcmp}) is truncated in practice, $n=1\ldots N$.
The truncation $N$ is determined by the physics of the system
(ultimately the lattice action), the simulation parameters (most importantly the
lattice energy cutoff), and by the choice of operators. 
The matrix elements (\ref{Cphin}) can be interpreted as
components of $N$ $M$-dimensional vectors $\phi_n$,
where $M$ is the size of the correlation matrix $C(t,t_0)$.
Following L{\"u}scher and Wolff \cite{Luscher:1990ck} we diagonalize the
correlation matrix separately on each time slice, say
\begin{subequations}
\begin{eqnarray}
\sum_{j=1}^{M} C_{ij}(t,t_0) v_{mj}(t,t_0) &=& C_m(t,t_0) v_{mi}(t,t_0) \label{Cv} \\
\sum_{i=1}^M v_{mi}^\ast(t,t_0) v_{m^\prime i}(t,t_0) &=& \delta_{m m^\prime}\,, \label{vv}
\end{eqnarray}
\end{subequations}
denoting the eigenvectors by $v_m$ and the eigenvalues by $C_m$, $m=1\ldots M$.
Now assume the following:
\begin{enumerate}
\item The energies $E_n$ in (\ref{Cdcmp}) are non degenerate,
and ordered $E_1<E_2<\ldots<E_N$.
\item The vectors $\phi_n$ are linearly independent, this implies $N\leq M$.
\item There is a $t_C$ such that for all $t\geq t_C$ the eigenvalues are ordered
$C_1\geq C_2\geq \ldots \geq C_M$. 
\end{enumerate}
Under those conditions a theorem proven in \cite{Luscher:1990ck} states that
for all $n=1\ldots N$
\begin{equation}
\lim_{t\rightarrow\infty}C_n(t,t_0) = Z_n e^{-E_n(t-t_0)}
[1+O(e^{-\Delta E_n(t-t_0)})]\,,
\label{Lemma}\end{equation}
where $Z_n>0$ and $\Delta E_n=\min_{n^\prime\neq n}\{|E_{n^\prime}-E_n|\}$
is the distance to the energy closest to $E_n$.
We are interested in the structure of the spectral representation of $C_n(t,t_0)$.
Toward this end, applying (\ref{vv}) to (\ref{Cv}) and then inserting (\ref{Cdcmp}),
one obtains
\begin{equation}
C_m(t,t_0) = \sum_{n\neq 0} |\sum_{i=1}^M v_{mi}^\ast(t,t_0) \phi_{ni}|^2
e^{-E_n(t-t_0)}\,.
\label{Cm}\end{equation}
In addition to items 1.--3. above we will also assume:
\begin{enumerate}\setcounter{enumi}{3}
\item In the large-$t$ limit the first $N^\prime$ of the eigenvectors of
(\ref{Cv},\ref{vv}) converge in the sense that in
\begin{equation}
\lim_{t\rightarrow\infty}v_{mi}(t,t_0) = \eta_m(t,t_0)\,v_{mi},
\quad m=1\ldots N^\prime\,,
\label{etav}\end{equation}
the vectors $v_m$ are constant, being multiplied by a $t$-dependent
phase factor, $|\eta_m(t,t_0)|=1$.
\end{enumerate}
Note that $N^\prime\leq N\leq M$.
Thus, using (\ref{etav}) and (\ref{Lemma}) in (\ref{Cm}) we arrive at
\begin{equation}
\sum_{n\neq 0} |\sum_{i=1}^M v_{mi}^\ast \phi_{ni}|^2 e^{-E_n(t-t_0)}
=Z_m e^{-E_m(t-t_0)}\,,
\label{Cinfty}\end{equation}
for $m=1\ldots N^\prime$.
Since all $E_n$ are different the exponentials are linearly independent
functions of $t$, hence
\begin{equation}
|\sum_{i=1}^M v_{mi}^\ast \phi_{ni}|^2 = Z_n\delta_{mn},\quad m=1\ldots N^\prime\,.
\end{equation}
The square root of this is
\begin{equation}
\sum_{i=1}^M v_{mi}^\ast \phi_{ni} = \zeta_m\sqrt{Z_m}\delta_{mn},\quad m=1\ldots N^\prime\,,
\label{vphiZ}\end{equation}
with $|\zeta_m|=1$. The eigenvectors of (\ref{Cv},\ref{vv}) satisfy a completeness relation
in $M$-dimensional space. We split it into two parts
\begin{equation}
\sum_{m=1}^{N^\prime} v_{mj}(t,t_0) v_{mi}^\ast(t,t_0) +
\sum_{k>N^\prime}^M v_{kj}(t,t_0) v_{ki}^\ast(t,t_0) = \delta_{ji}\,.
\label{split}\end{equation}
According to the assumption (\ref{etav}) all terms in the first sum will individually
converge in the large-$t$ limit. The individual terms in the second sum might not,
however, it must of course become $t$-independent as a whole,
\begin{equation}
\Pi_{ji}=\lim_{t\rightarrow\infty}\sum_{k>N^\prime}^M v_{kj}(t,t_0) v_{ki}^\ast(t,t_0)\,.
\label{defPi}\end{equation}
Clearly the projector $\Pi=\Pi^2=\Pi^\dagger$ is orthogonal on the space defined
by the span of the $v_m$,
\begin{equation}
\sum_{i=1}^M \Pi_{ji} v_{mi}=0,\quad{\rm for}\quad m=1\ldots N^\prime\,.
\label{splitPi}\end{equation}
Thus, as $t\rightarrow\infty$, equation (\ref{split}) assumes the form
\begin{equation}
\sum_{m=1}^{N^\prime} v_{mj} v_{mi}^\ast + \Pi_{ji} = \delta_{ji}\,.
\label{split2}\end{equation}
Finally, operating with $\sum_{m=1}^{N^\prime}v_{mj}\ldots$ on both sides of (\ref{vphiZ})
and then using (\ref{split2}) gives
\begin{equation}
\phi_{n}=\zeta_n\sqrt{Z_n}v_{n}+\Pi\phi_n,\quad{\rm for}\quad n=1\ldots N^\prime\,.
\label{PhiZPi}\end{equation}
This result relates
the matrix elements $\phi_{ni}=\langle 0|\hat{\Phi}_i^\dagger(t_0)|n\rangle$
to the solutions of the $t$-dependent eigenvalue problem (\ref{Cv}, \ref{vv})
in the large-$t$ limit.
An immediate consequence, derived by taking the square of (\ref{PhiZPi}), is
$\phi_n^\dagger\phi_n=Z_n+\phi_n^\dagger\Pi\phi_n$, or
\begin{equation}
\sum_{i=1}^M |\langle n|\Phi_i(t_0)|0\rangle|^2 = Z_n + |\!|\Pi\phi_n|\!|^2,
\quad{\rm for}\quad n=1\ldots N^\prime\,.
\label{Zn2}\end{equation}
Thus $Z_n$ is a lower bound on the total probability for 
(incoherent) excitations by a set of operators $\Phi_i, i=1\ldots M$,
into a certain state $|n\rangle$.
In principle the value of $N^\prime$ can be computed from (\ref{etav}).
In practice this is hard to accomplish, because the components
of the eigenvectors fluctuate strongly.
If the set of $M$ operators couples to all $N$ available physical states
it seems reasonable to expect that $N^\prime=N$.
In case that $N<M$ we have used more operators than quantum states are available
in the system. Then the projector term compensates for over counting the
physical degrees of freedom.

Aside from (\ref{Zn2}) there is an alternative way of interpreting the $Z_n$.
Based on (\ref{etav}) define the meson-meson fields
\begin{equation}
\Psi_m(t)=\sum_{i=1}^M v_{mi}\Phi_i(t)
\quad{\rm for}\quad m=1\ldots N^\prime\,,
\label{Psim}\end{equation}
and consider the $N^\prime\times N^\prime$ correlator matrix
\begin{equation}
D_{mm^\prime}(t,t_0)=\langle\hat{\Psi}_{m}^\dagger(t)\hat{\Psi}_{m^\prime}(t_0)\rangle\,.
\label{D22}\end{equation}
Inserting (\ref{Psim}) into (\ref{D22}), then using (\ref{Cdcmp}) and (\ref{vphiZ}),
it is straightforward to show that
\begin{equation}
D_{mm^\prime}(t,t_0)=\delta_{mm^\prime} Z_m e^{-E_m(t-t_0)}\,.
\label{Dmm}\end{equation}
On the other hand, starting from (\ref{D22}), the diagonal elements have
the standard decomposition
\begin{equation}
D_{mm}(t,t_0)=\sum_{n\neq 0}|\langle n|\Psi_m(t_0)|0\rangle|^2 e^{-E_n(t-t_0)}\,.
\label{Dpsin}\end{equation}
Comparison of (\ref{Dmm}) and (\ref{Dpsin}), using linear independence
of the exponentials again, then gives
\begin{equation}
|\langle n|\Psi_m(t_0)|0\rangle|^2=\delta_{nm}Z_n
\quad{\rm for}\quad n,m=1\ldots N^\prime\,.
\label{Zn2a}\end{equation}
Thus $Z_n=|\langle n|\Psi_n(t_0)|0\rangle|^2$ is the excitation probability
of the state $|n\rangle$ due to an operator $\Psi_n(t)$ that is optimal, with regard
to $|n\rangle$, within the linear space of the original set $\Phi_i(t), i=1\ldots M$.

\section{\label{sec:results}Discussion of results}

The significance of the above is that it suggests an analysis
strategy for the spectral features of the two-meson system:
Diagonalize the time correlation matrix (\ref{Cij}) on each time slice,
\begin{eqnarray}
\left(
\begin{array}{cc}
C_{11}(t,t_0) & C_{12}(t,t_0) \\
C_{21}(t,t_0) & C_{22}(t,t_0)
\end{array}
\right)
\left(
\begin{array}{c}
v_{m1}(t,t_0) \\
v_{m2}(t,t_0)
\end{array}
\right)
&=& \nonumber\\
&&\hspace{-22ex}
C_m(t,t_0)
\left(
\begin{array}{c}
v_{m1}(t,t_0) \\
v_{m2}(t,t_0)
\end{array}
\right),
\label{Ce}\end{eqnarray}
$m=1,2$. Then view each eigenvalue as a separate time correlation function
$C_m(t,t_0)$ subject to spectral analysis. In particular, seek to extract a spectral
representation of the form $\int d\omega\,\rho(\omega)\,e^{-\omega(t-t_0)}$,
see (\ref{Fc}). This can be done by Bayesian inference \cite{Jar96,Box73,Fiebig:2002sp}.
The expected structure of
the spectral density $\rho$ is a linear combination of peaks, see (\ref{rhodelta}).
In an actual numerical simulation those will have finite widths. More
importantly, because the main results of Sec.~\ref{sec:analysis},
including (\ref{Zn2}) and (\ref{Zn2a}), require the limit $t\rightarrow\infty$
only the lowest-energy peak of each $C_m(t,t_0)$
is significant. Physical information that can be extracted from the peak
includes the energy $E_n$ of the state $|n\rangle$, and the strength $Z_n$ of
excitations by means of the set of operators employed.

\begin{figure}
\includegraphics[angle=0,width=42mm]{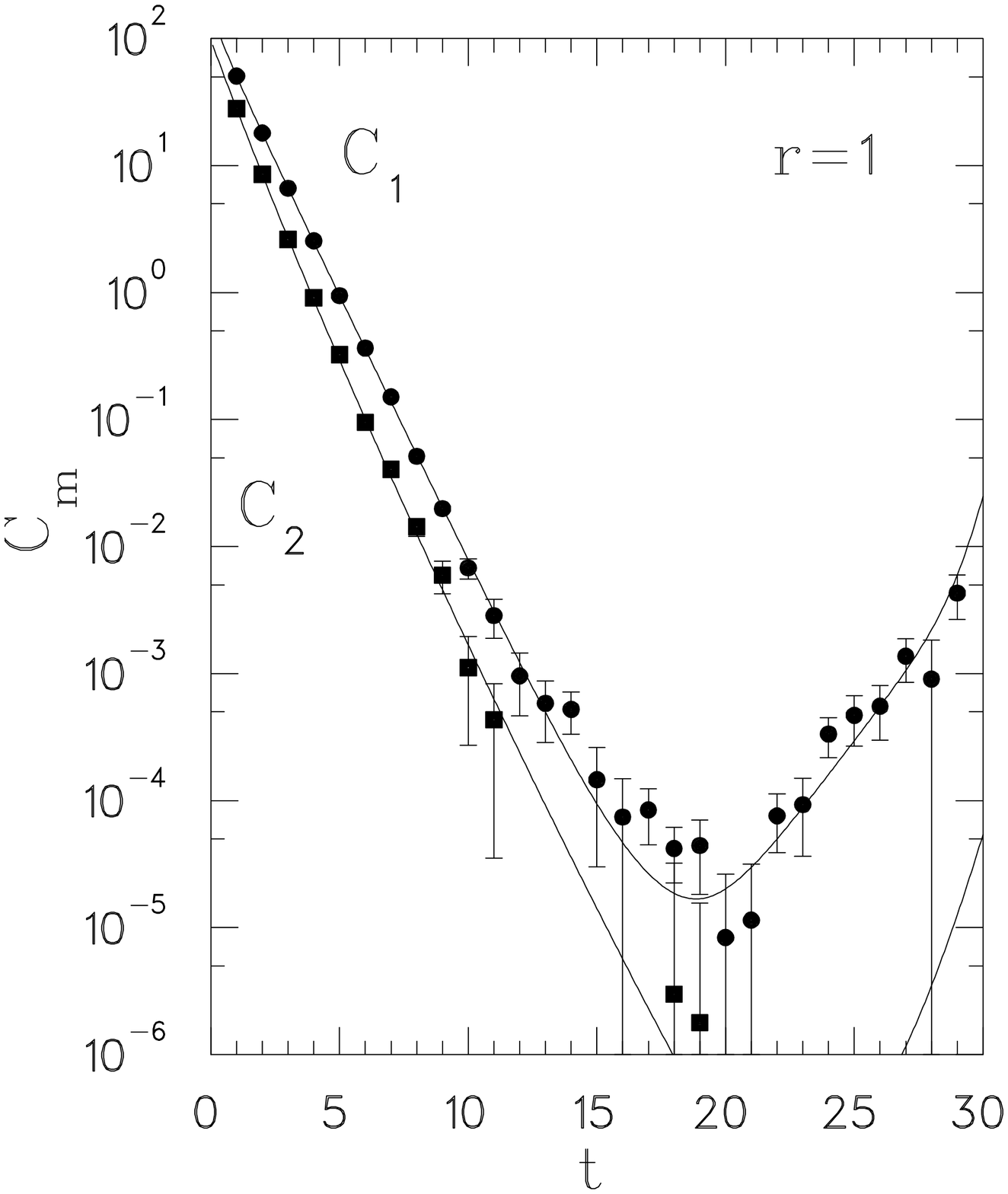}
\includegraphics[angle=0,width=42mm]{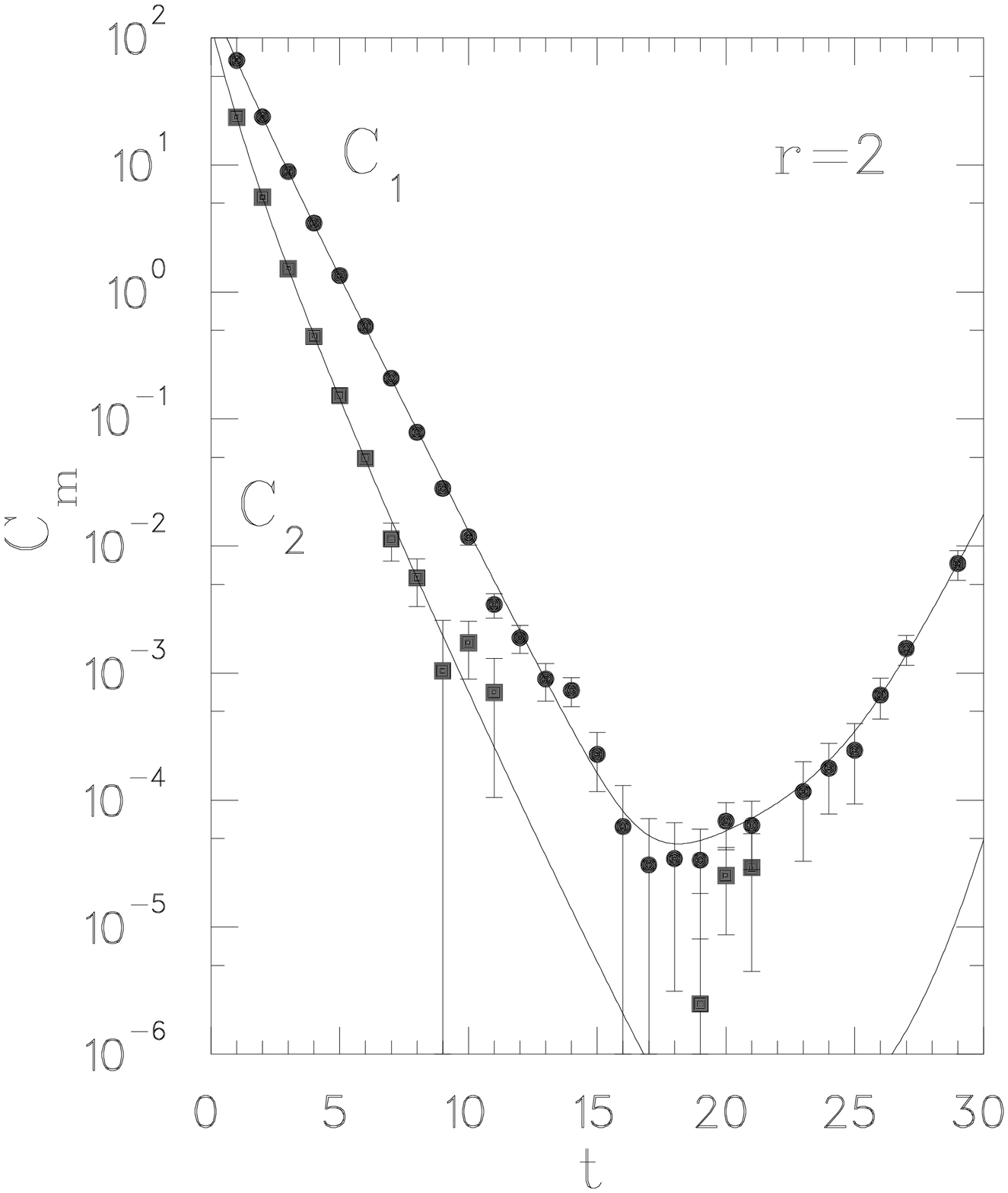}\vspace{2ex}\\
\includegraphics[angle=0,width=42mm]{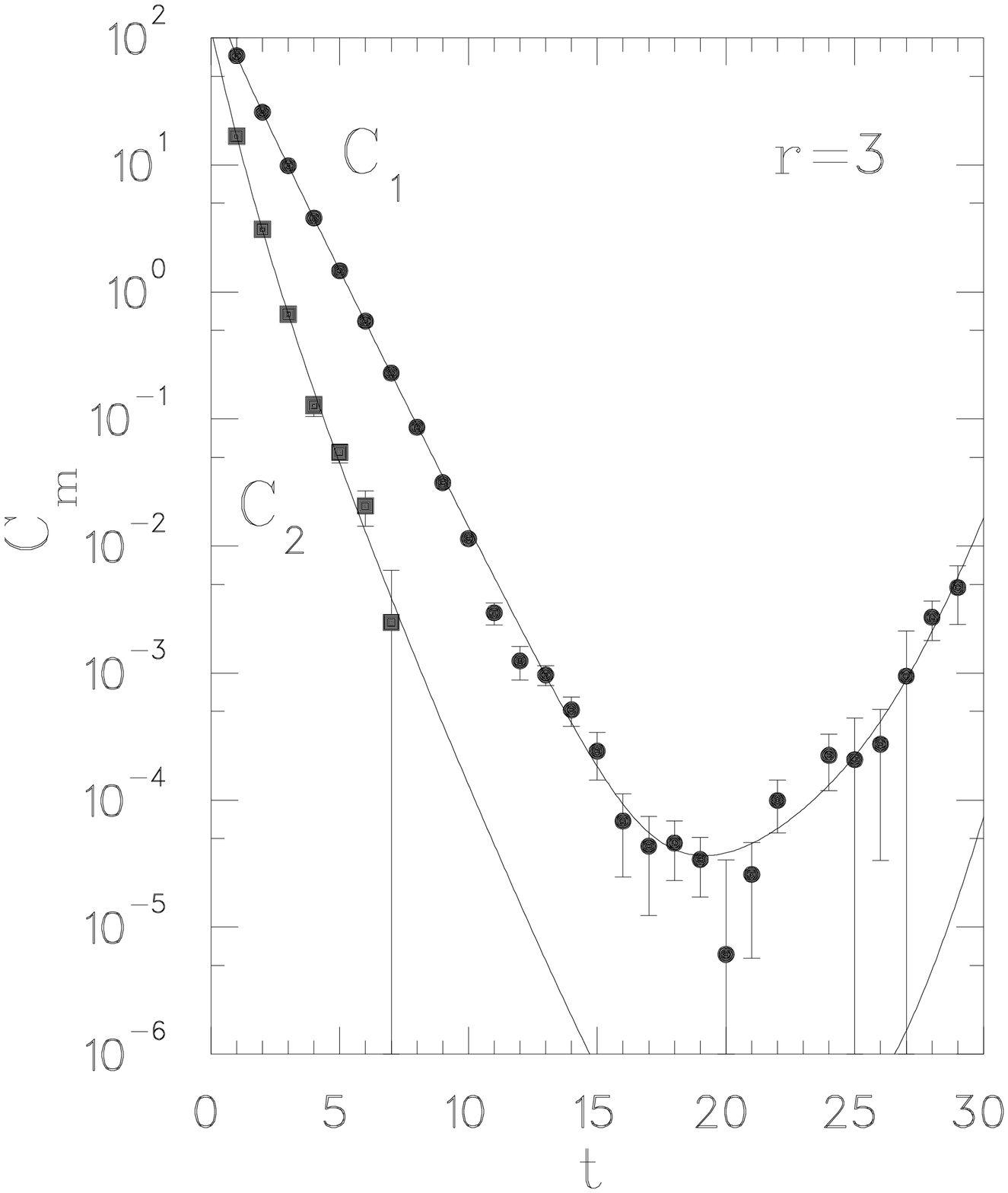}
\includegraphics[angle=0,width=42mm]{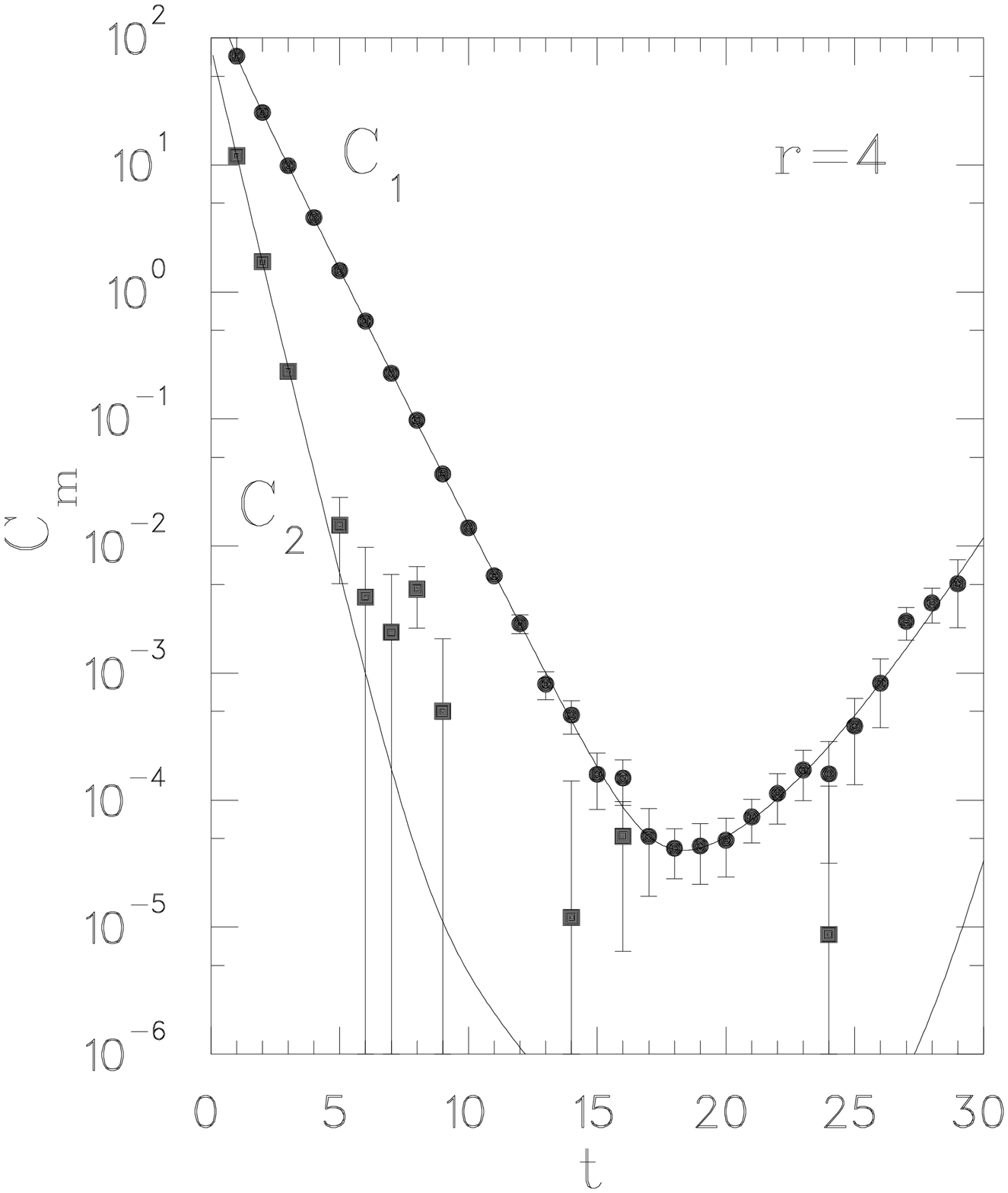}
\caption{\label{fig2}Eigenvalue correlation functions at meson-meson relative
distances $r=1,2,3,4$. We show $C_1(t,t_0)$ as filled circles, and $C_2(t,t_0)$
as filled squares. Indicated are statistical errors from $N_U=708$ gauge configurations.
A missing plot symbol means that the statistical error exceeds the value of a data point.
The lines are plots of the model (\ref{Fd}) using the Bayesian results for the
spectral densities $\rho$ from Figs.~\protect\ref{fig3} and \protect\ref{fig4}.}
\end{figure}

The eigenvalue correlators $C_m(t,t_0), m=1,2$, of (\ref{Ce}) are displayed in
Fig.~\ref{fig2} for four meson-meson relative distances $r=1\ldots 4$. 
Due to the non-local structure of the two-meson operators,
leading to loop-loop correlations, somewhat noisy data are unavoidable.
Link variable fuzzing and operator smearing was used to 
be able to use `earlier' time slices, recall Sect.~\ref{sec:operators}.
Not surprisingly, the lower eigenvalue $C_2(t,t_0)$ is more susceptible
to noise than $C_1(t,t_0)$.
Backward going propagation is present, though suppressed by four to five orders
of magnitude.\footnote{Note that (anti)periodic boundary conditions in the
lattice action do not in general translate into similar behavior of the correlation
functions. Here, these are some combination of `$\cosh$' and `$\exp$' behavior
akin to local meson and loop like operators.}
By inspection of Fig.~\ref{fig2} it is apparent that the condition
$C_1(t,t_0) \geq C_2(t,t_0)$ is fulfilled for $t\geq t_0$.
This is directly obvious for most of the forward temporal range, say $0<t<20$,
and for $20\leq t<30$ it can be inferred from the global fits to $C_2(t,t_0)$.
Indicative of a violation would be the possibility of
being able to smoothly connect two sets of consecutive data points in such a way
that the two resulting curves would intersect.\footnote{For example
$c_1(t,t_0)=a_1 e^{-b_1(t-t_0)}$ and
$c_2(t,t_0)=a_2 e^{-b_2(t-t_0)}$ intersect at $t_C-t_0=\ln(a_2/a_1)/(b_2-b_1)$.
The observation simply means that $t_C\leq t_0$.}
We thus observe that the above ordering of eigenvalues, as stated in
Sec.~\ref{sec:analysis}, is satisfied.

\begin{figure}
\includegraphics[angle=0,width=42mm]{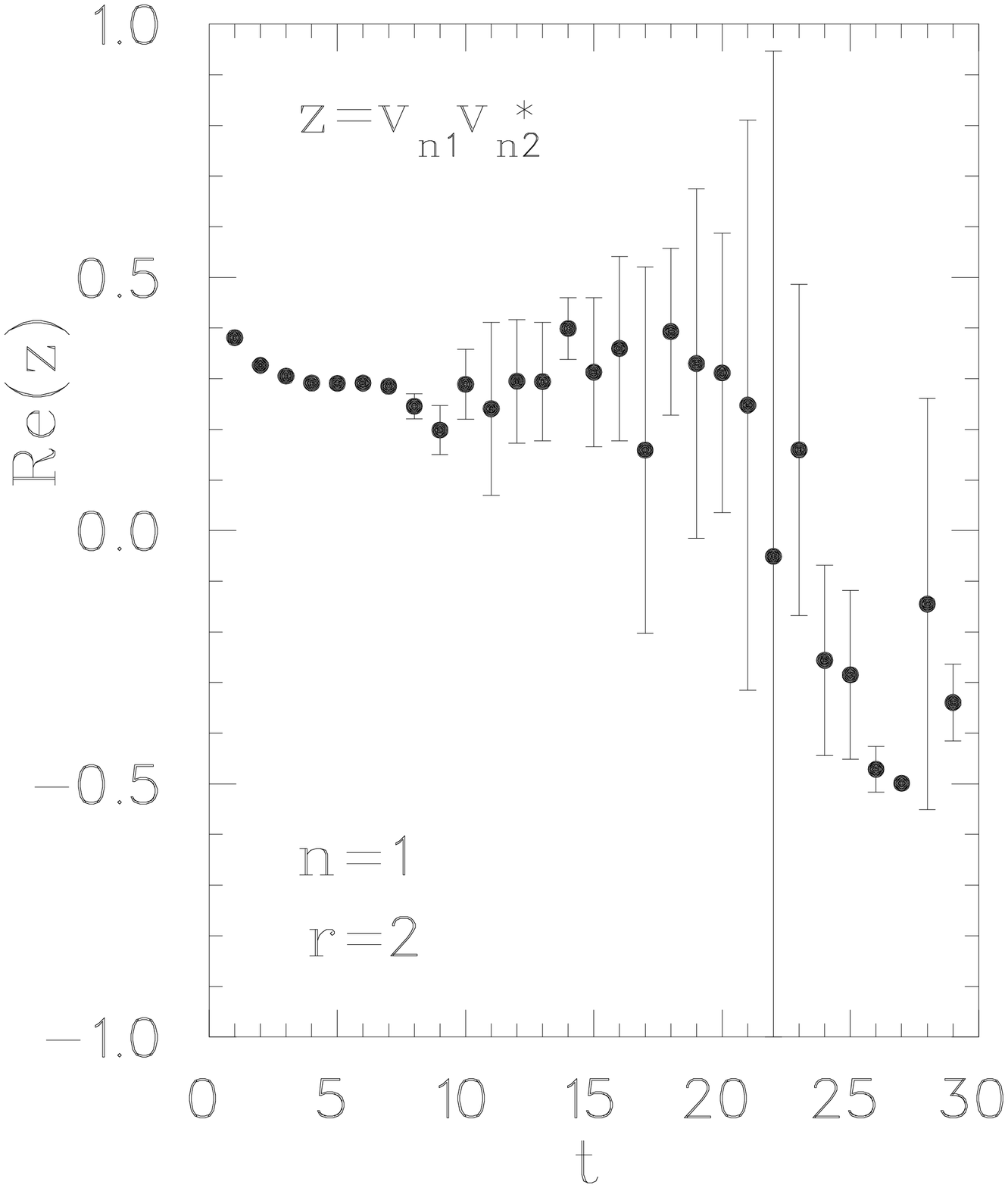}
\includegraphics[angle=0,width=42mm]{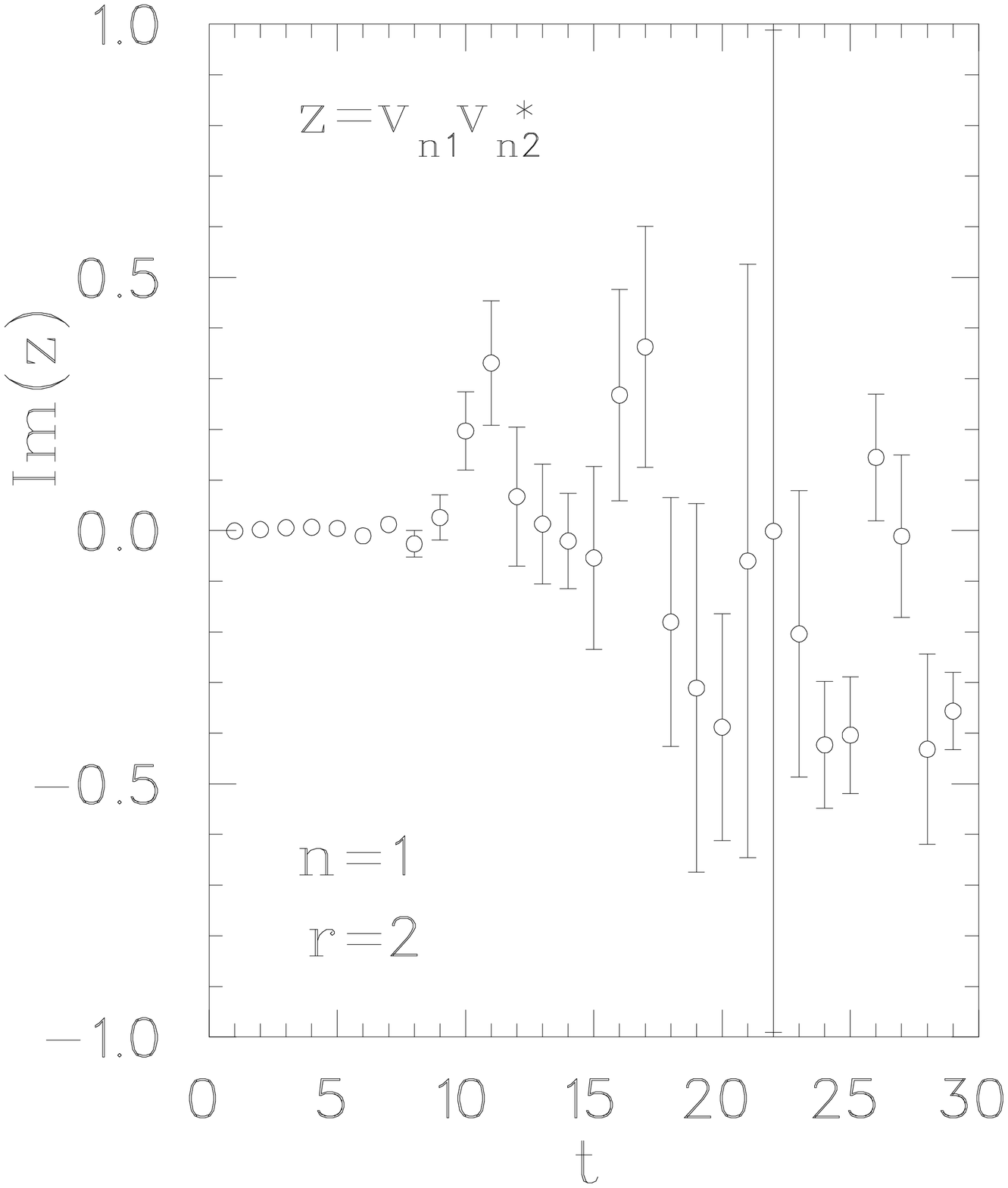}
\caption{\label{fig10gh} Real and imaginary parts of
$z=v^{\ast}_{n1}(t,t_0)v_{n2}(t,t_0)$, the example is for the ground
state $n=1$ and relative distance $r=2$.
The numerical constancy of $z$ within errors (excluding the backward propagation
in the region $t\agt 20$) illustrates the validity of (\protect\ref{etav}).}
\end{figure}

For asymptotic times individual components of the eigenvectors
in (\ref{Ce}) show large statistical fluctuations.
In Fig.~\ref{fig10gh} a typical example of the (complex) values of
$z=v_{n1}(t,t_0)v^{\ast}_{n2}(t,t_0)$ is displayed.
This quantity is relevant to the assumption (\ref{etav}).
In the region of forward propagation, say $t\alt 20$, the real part of $z$ exhibits
increasing stochastic errors as $t$ becomes large, but stabilizes within those
bounds. A similar statement can be made for the imaginary part of $z$, adding that
it is consistent with zero. This justifies (\ref{etav}) within the limitations given
by the quality of the
data.\footnote{Note that (\protect\ref{etav}) is stated for the forward
propagation region of the correlation functions. It may have to be modified for
backward propagation.}

\begin{figure}
\includegraphics[angle=0,width=42mm]{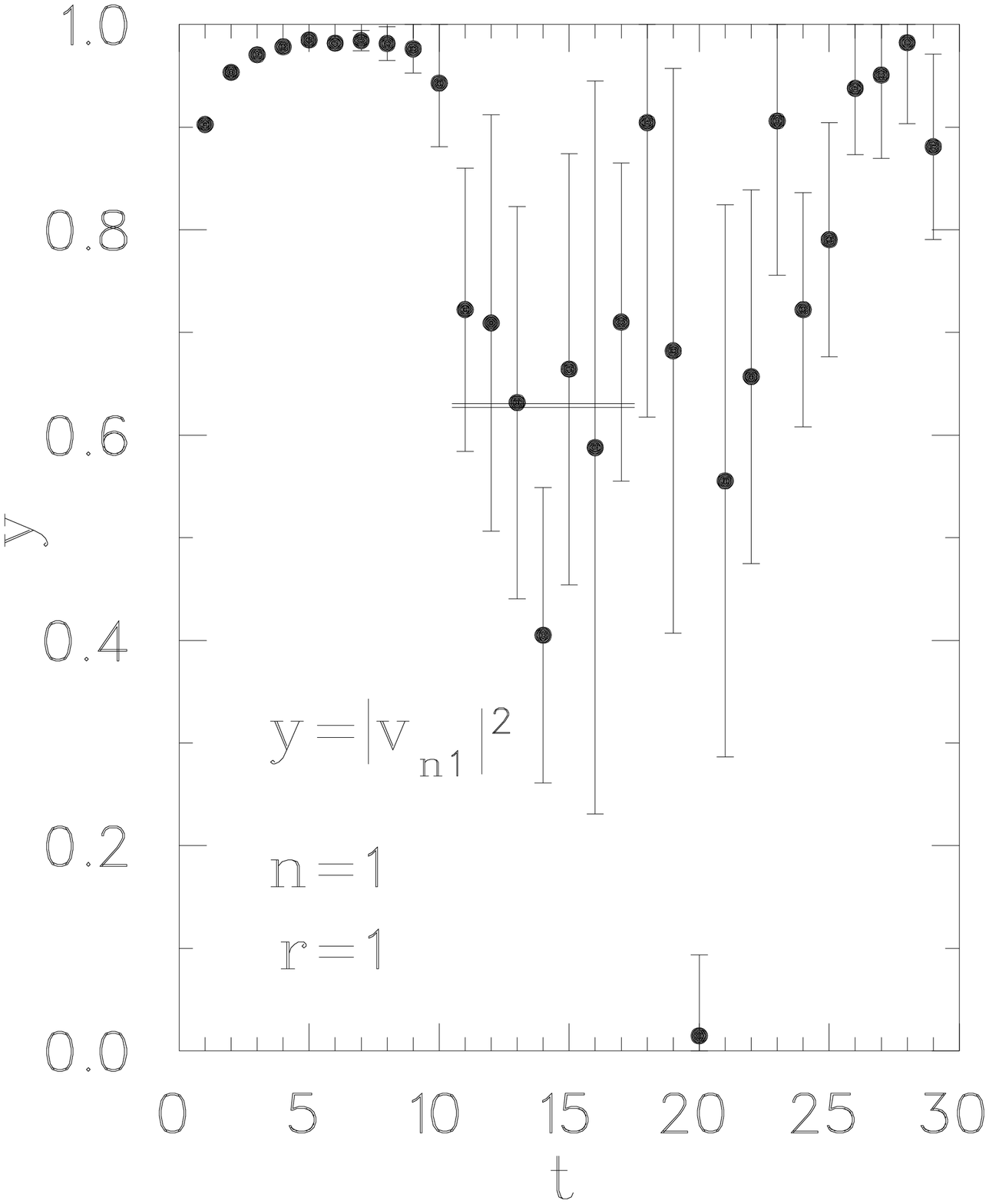}
\includegraphics[angle=0,width=42mm]{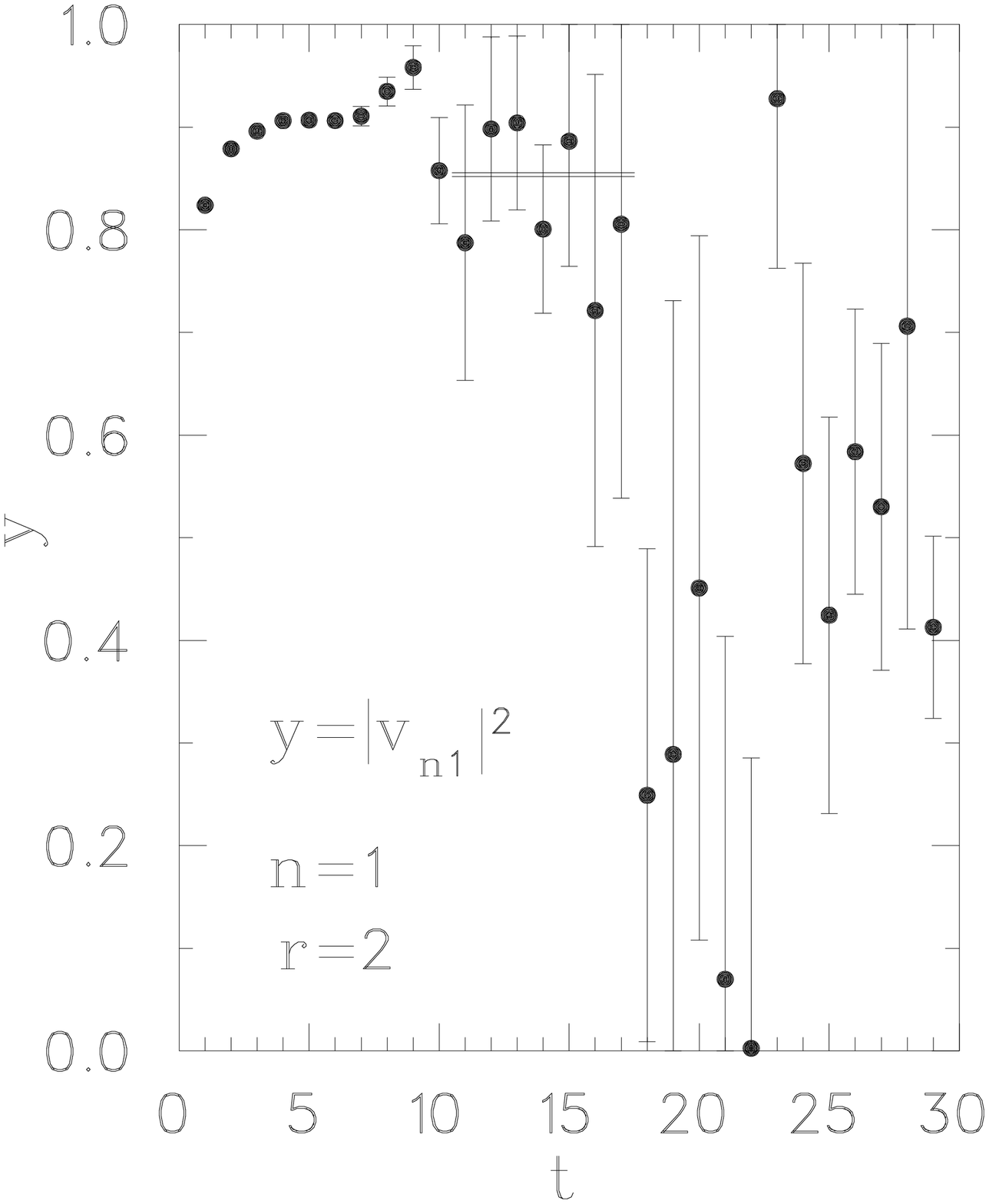}\vspace{2ex}\\
\includegraphics[angle=0,width=42mm]{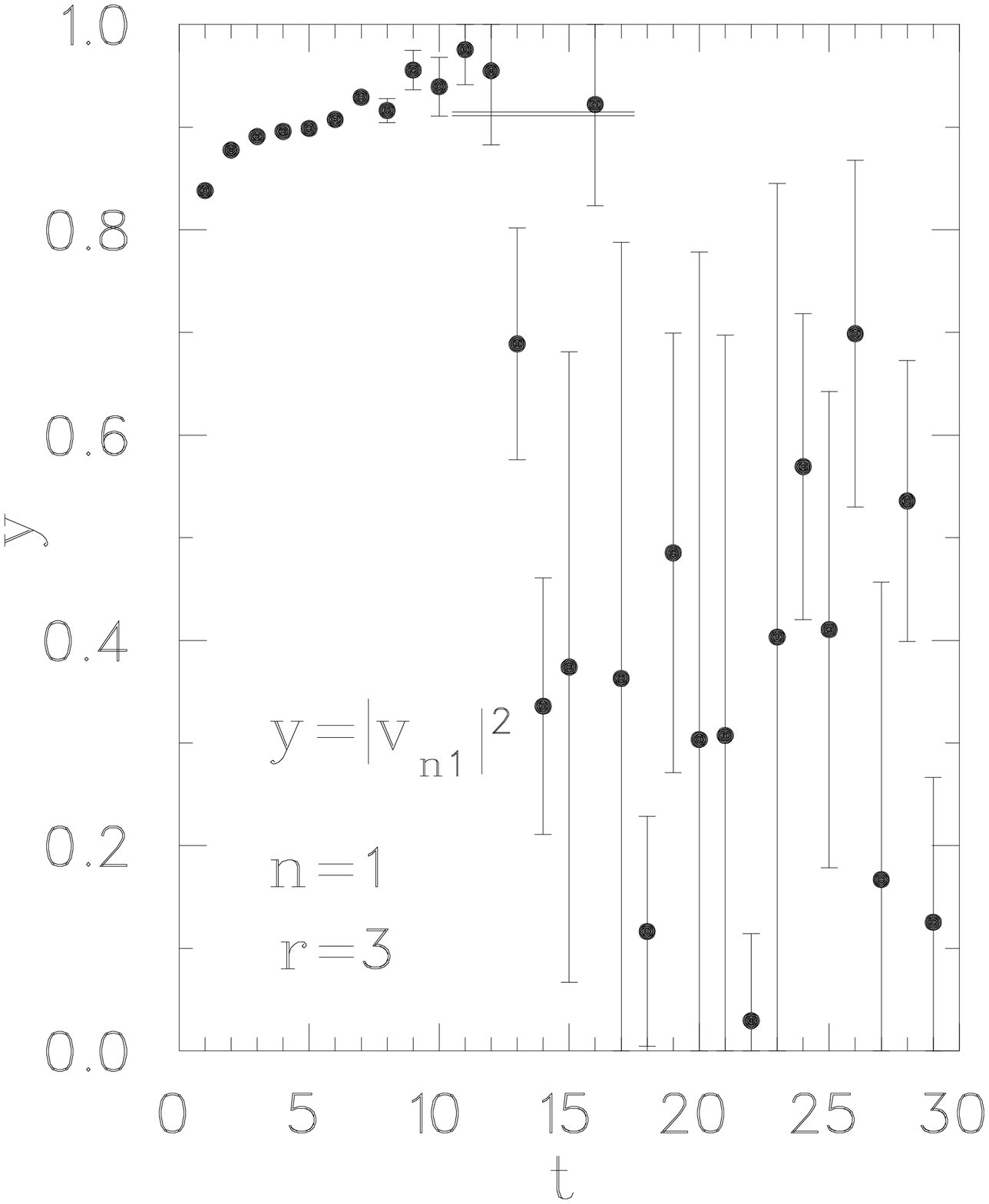}
\includegraphics[angle=0,width=42mm]{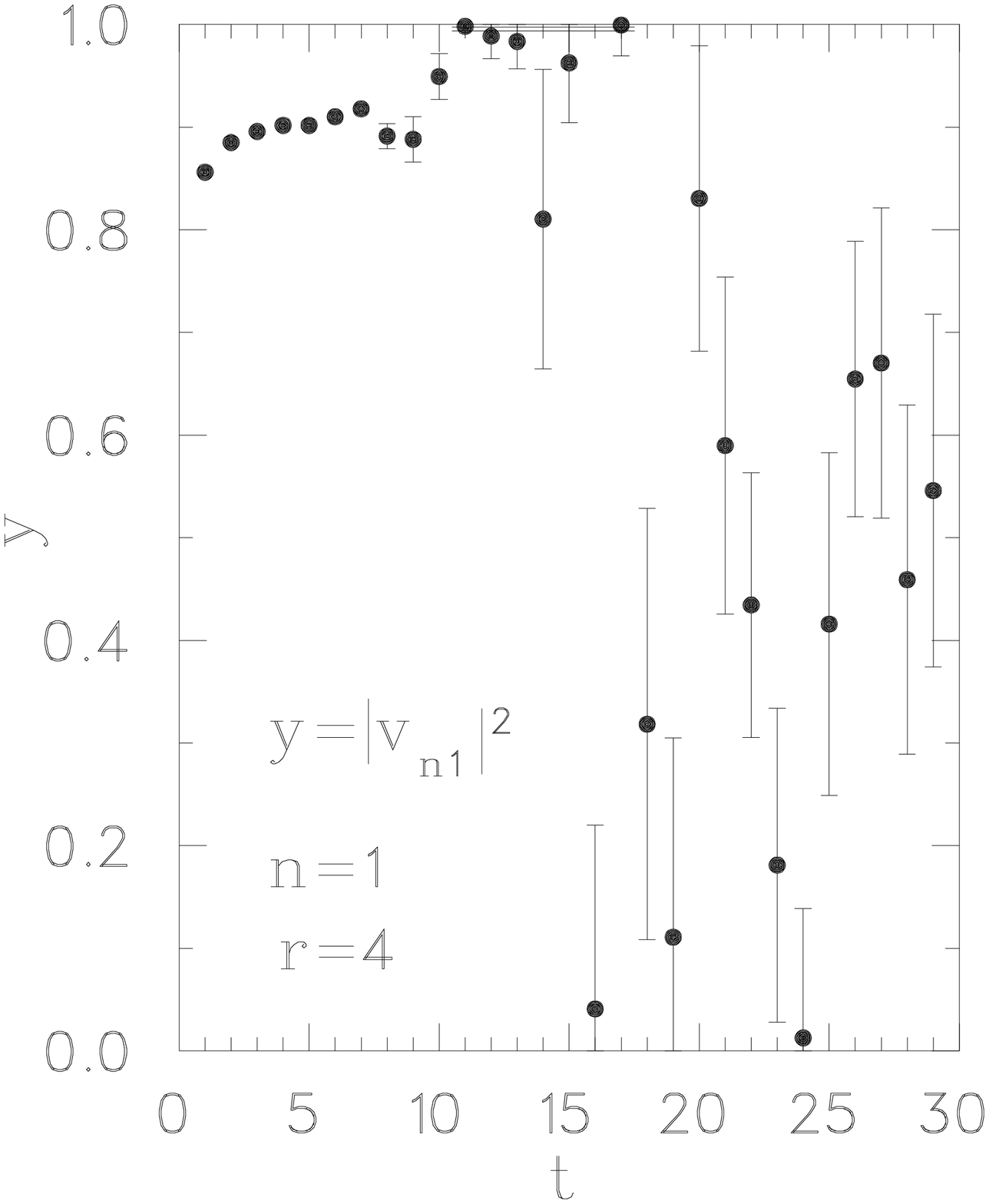}
\caption{\label{fig10abcd} Time dependence of the squared magnitudes
$y=|v_{mi}(t,t_0)|^2$ of the $i=1$ component of the ground state $m=1$,
for relative distances $r=1,2,3,4$. The uncertainties are jackknife
standard errors. The horizontal lines indicate constant model fits in
the range $11\leq t\leq 17$, $s$ from Tab~\protect\ref{tab:sfit}.} 
\end{figure}

In Fig.~\ref{fig10abcd} we show the magnitudes $|v_{11}(t,t_0)|^2$ of eigenvector
components versus $t$ for four relative distances.
Substantial noise notwithstanding we have strictly applied the
prescription of Sect.~\ref{sec:analysis} and computed
the eigenvalue correlators following (\ref{Cv},\ref{vv}) verbatim 
\begin{equation}
C_m(t,t_0)=\sum_{i,j=1}^{2}v_{mi}^\ast(t,t_0) C_{ij}(t,t_0) v_{mj}(t,t_0)\,,
\label{vCv}\end{equation}
for $m=1,2$. The corresponding spectral analysis is discussed below.

Stochastic fluctuations of the eigenvector components are enhanced by
diagonalizing (\ref{Ce}) separately on each time slice.
An attractive alternative is to replace the eigenvector components
$v_{mi}(t,t_0)$ in (\ref{vCv}) with time averaged components $\bar{v}_{mi}$
taken at asymptotic times. Within the present data
set the available time window is $10\alt t\alt 20$, excluding backward propagation.
We have made fits to $|v_{11}(t,t_0)|^2$ with a constant model $s$ in the time
slice range $11\le t \le 17$, and also with a control set in $11\le t \le 19$.
The results are listed in Tab.~\ref{tab:sfit}.
Specific values for $s$ give rise to, time independent, vectors
\begin{equation}
\bar{v}_1=\left(\begin{array}{c}\sqrt{s} \\  \sqrt{1-s}\end{array}\right) \quad
\bar{v}_2=\left(\begin{array}{c}\sqrt{1-s} \\ -\sqrt{s}\end{array}\right)\,,
\label{w3w4}\end{equation}
up to an arbitrary phase. Thus we have also performed a spectral analysis based
on the correlators
\begin{equation}
\bar{C}_m(t,t_0)=\sum_{i,j=1}^{2}\bar{v}_{mi}^\ast C_{ij}(t,t_0) \bar{v}_{mj}\,,
\label{bCb}\end{equation}
for $m=1,2$. Besides smoothing out statistical fluctuations of the eigenvector
components the advantage of (\ref{bCb}) is that the asymptotic form of the
correlator (\ref{vCv}) is now used on all time slices. This should improve the
signal derived from the Bayesian spectral analysis which makes use of data
on all times slices.
\begin{table}
\caption{\label{tab:sfit}Time averages $s$ of $|v_{11}(t,t_0)|^2$ and
the corresponding variances $\Delta s$. Fits for two time slice ranges are listed.}
\begin{ruledtabular}
\begin{tabular}{ccccccc}
& & \multicolumn{2}{c}{$11\leq t\leq 17$} & & \multicolumn{2}{c}{$11\leq t\leq 19$} \\
\hline
$r$ & & $s$ & $\Delta s$ & & $s$ & $\Delta s$ \\
\colrule
1.0 & & 0.629 & 0.132 & & 0.645 & 0.138 \\
2.0 & & 0.854 & 0.060 & & 0.831 & 0.134 \\
3.0 & & 0.913 & 0.173 & & 0.861 & 0.259 \\
4.0 & & 0.996 & 0.033 & & 0.994 & 0.046 \\
\end{tabular}
\end{ruledtabular}
\end{table}

\begin{figure}
\includegraphics[angle=0,width=42mm]{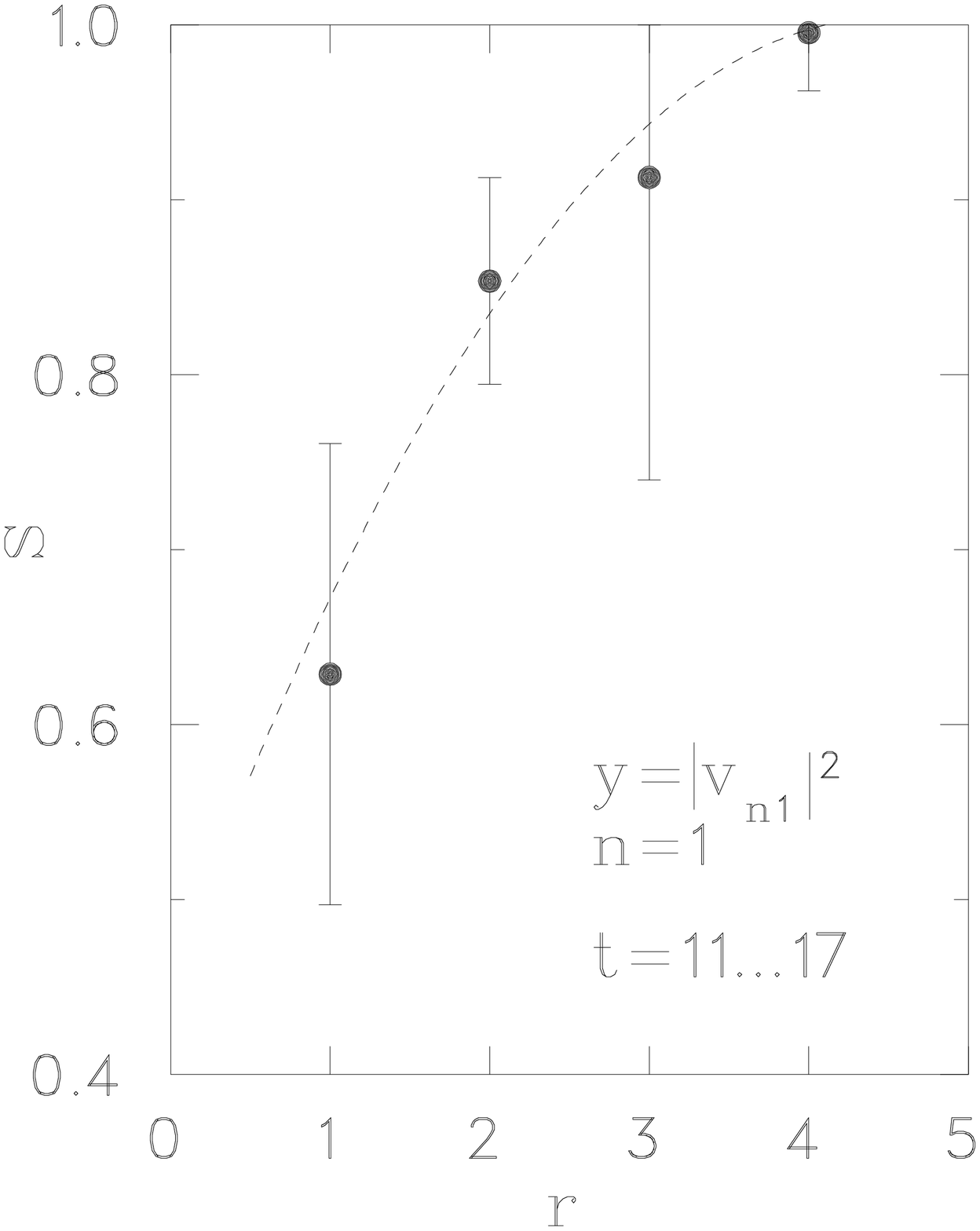}
\includegraphics[angle=0,width=42mm]{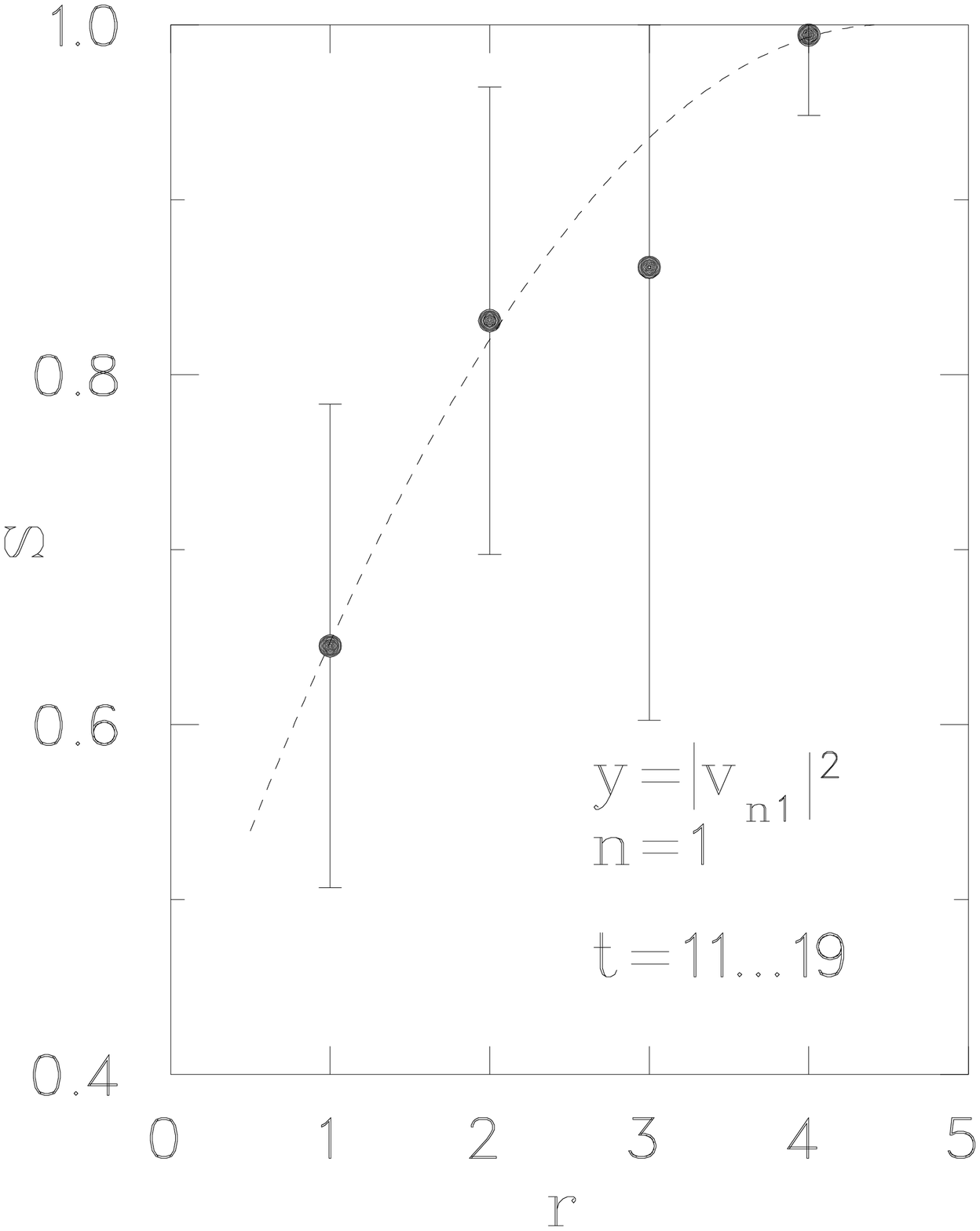}
\caption{\label{fig11R} Asymptotic time averaged values $s$ of $|v_{11}(t,t_0)|^2$,
and their variances,
as functions of the relative meson-meson distance $r$. Results for two time
slice ranges are shown. The dashed curves are quadratic polynomial fits.}
\end{figure}

The values of $s$ inform us about operator mixing as the relative distance $r$ changes.
They are a measure of how strongly the operator $\Phi_1$ couples
to a meson-meson system in the ground state. As Fig.~\ref{fig11R} shows this measure
distinctly decreases from about $1.0$ as $r$ becomes smaller.
Since $\Phi_1$ is designed to test quark exchange degrees of freedom we see
that those gradually become less important at smaller relative distances.
By the same token $1-s$ measures the ground state coupling strength of $\Phi_2$.
Thus we learn that
quark exchange is the dominant interaction mechanism at large $r$, while
gluon exchange gradually takes over as $r$ decreases. A glance at Fig.~\ref{fig11R}
reveals that the mechanisms become balanced ($s\approx 0.5$) at distances
$r$ somewhat less than $0.5$, or $0.25{\rm fm}$ in physical units. 

The Bayesian analysis of time correlation functions has been extensively discussed
in \cite{Fiebig:2002sp}, using the same lattice (raw) data. We here
briefly state the main points for coherence of presentation, but
otherwise refer the reader to \cite{Fiebig:2002sp}.
We expect the lattice data to fit the model (\ref{Fc}),
or rather its discretized form
\begin{equation}
F(\rho|t,t_0) \simeq \sum_{k=K_-}^{K_+}\rho_k e^{-\omega_k(t-t_0)}
\label{Fd}\end{equation}
with $\rho_k=\Delta\omega \rho(\omega_k)$ and $\omega_k=\Delta\omega k$.\footnote{
In Ref.~\protect\cite{Fiebig:2002sp} backward ($\omega<0$) and forward ($\omega>0$)
going exponentials were
normalized differently, giving rise to a modified spectral function $\rho_T$.
However, $\rho_T(\omega)=\rho(\omega)$ for $\omega\geq 0$.
Since we here only present $\omega\geq 0$ data there is no need to
distinguish between $\rho_T$ and $\rho$.}
The objective is to compute the spectral density function $\rho$.
Toward this end consider the functional
\begin{equation}
W[\rho]=\frac12\chi^2[\rho]-\alpha {\cal S}[\rho]\,.
\label{Wrho}\end{equation}
The first term involves the usual $\chi^2$-distance between the lattice data $C_m(t,t_0)$
and the model $F(\rho|t,t_0)$, computed with the full covariance matrix
derived from gauge field configuration statistics, and
\begin{equation}
{\cal S}[\rho] = \sum_{k=K_-}^{K_+} \left( \rho_k-m_k-\rho_k\ln\frac{\rho_k}{m_k}\right)\,,
\label{Smem}\end{equation}
is the (Shannon-Jaynes) entropy \cite{Sha49,Jay57b}. It plays
the role of a Bayesian prior \cite{Jar96}.
The configuration $m=\{m_k : K_- \leq k \leq K_+\}$ is called the
default model. Another parameter in (\ref{Wrho}) is the entropy strength $\alpha$.
The optimization problem $\chi^2[\rho]=\min$ is ambiguous because in practice
a reasonable resolution $\Delta\omega$ will result in the number $K_+-K_-+1$ of fit
parameters $\rho_k$ being much larger than the number $T$ of simulation data.
However, $W[\rho\/]$ has a unique absolute minimum \cite{Jar96}.
From the viewpoint of Bayesian statistics $\rho$ is interpreted as a random variable subject
to a certain probability distribution (posterior probability).
The most likely $\rho$ is the one that minimizes $W[\rho\/]$. 
Finding the minimum within this framework
is known as the maximum entropy method (MEM).
It is designed to minimize the {\sl information} not supported by the data.
Loosely speaking, we seek to minimize $\chi^2[\rho]$ while assuming
as little information as possible about the spectral density $\rho$.
To solve the optimization problem $W[\rho\/]=\min$, probabilistic methods
seem closest in spirit to the Bayesian stochastic interpretation of $\rho$.
We have thus employed simulated annealing \cite{Kir84}, or cooling, based
on the partition function
\begin{equation}
Z_W=\int [d\rho\/] e^{-\beta_W W[\rho\/]}\,.
\label{Zmem}\end{equation}
In \cite{Fiebig:2002sp} we have studied the dependence of the resulting spectral density
on (i) the entropy strength parameter $\alpha$, (ii) the default model $m$, and (iii)
the annealing start configuration. It was found that $\rho$ was essentially
independent of $\alpha$ and of $m$, and that the expectation values of low $\omega$
moments were independent
of the annealing start within errors native to the lattice data set.

\begin{figure}
\includegraphics[angle=0,width=41.5mm]{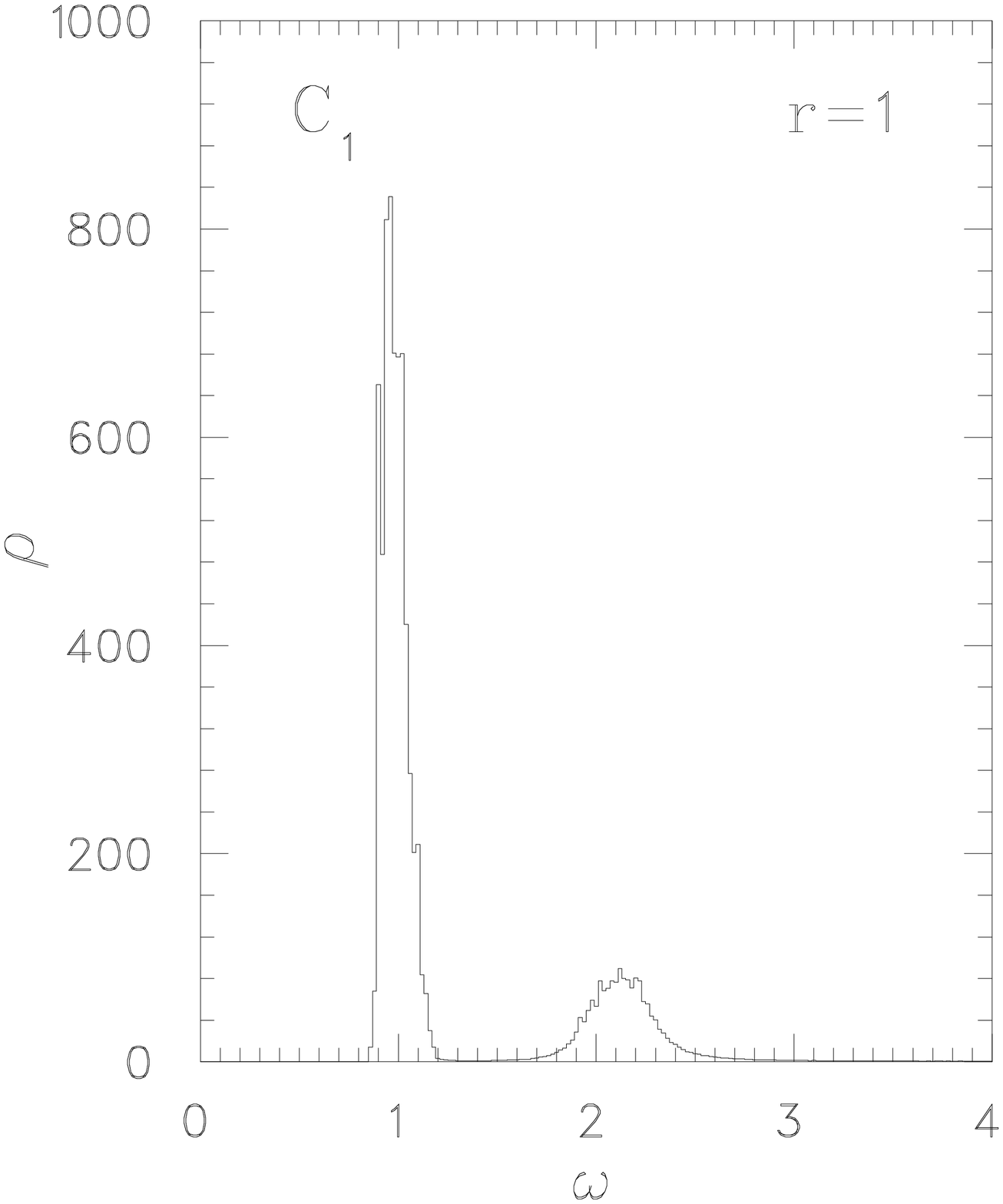}
\includegraphics[angle=0,width=41.5mm]{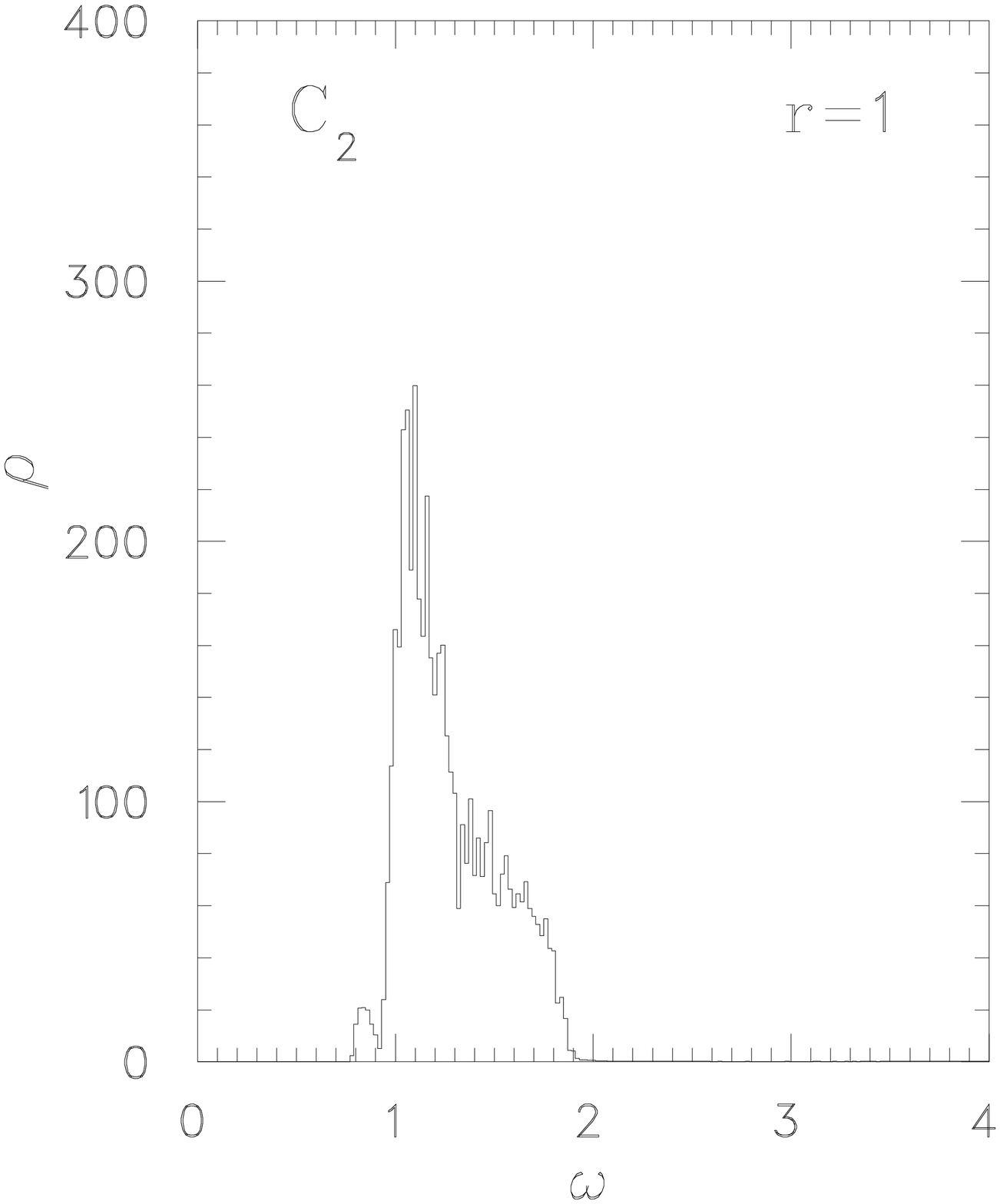}\vspace{2ex}\\
\includegraphics[angle=0,width=41.5mm]{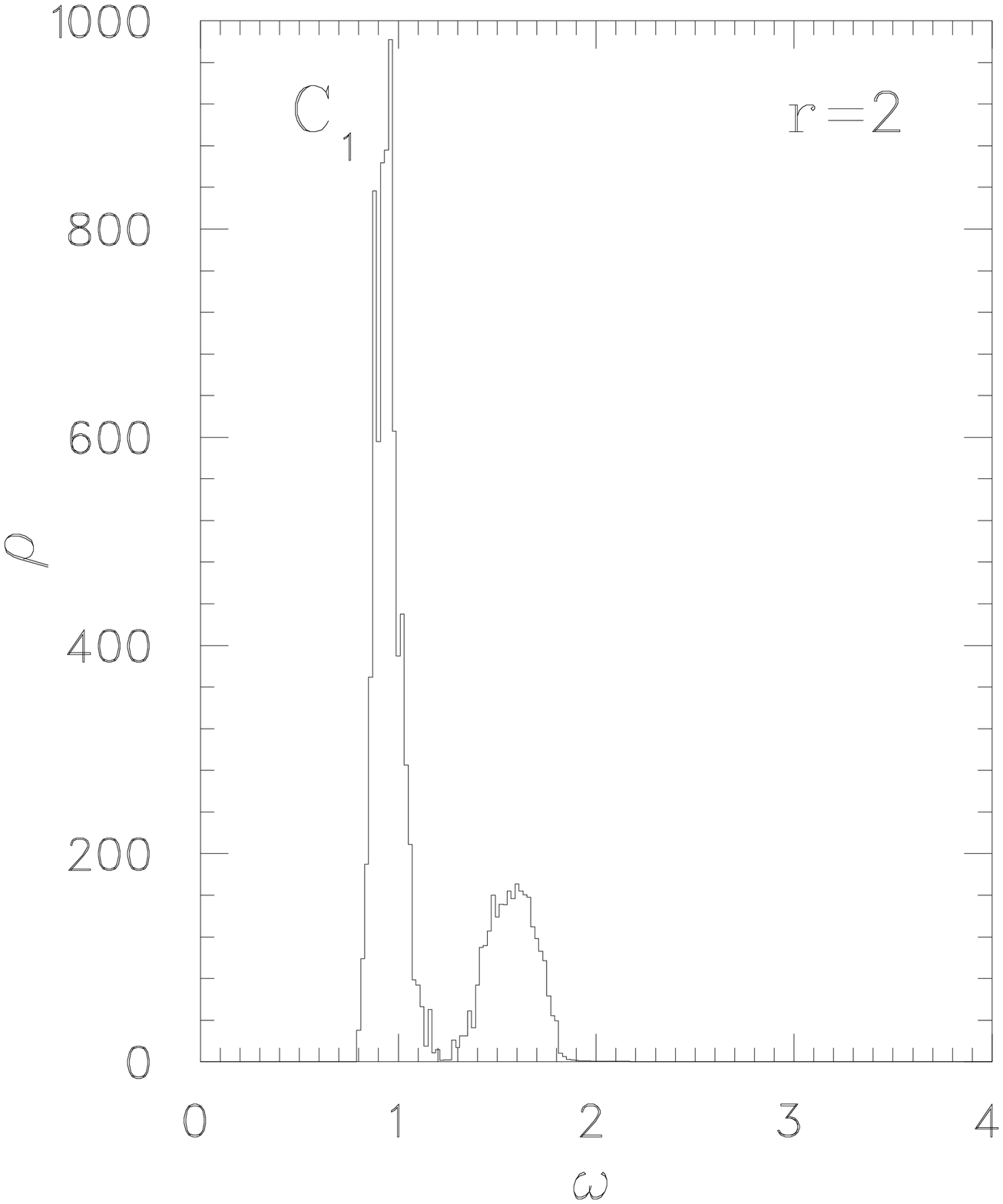}
\includegraphics[angle=0,width=41.5mm]{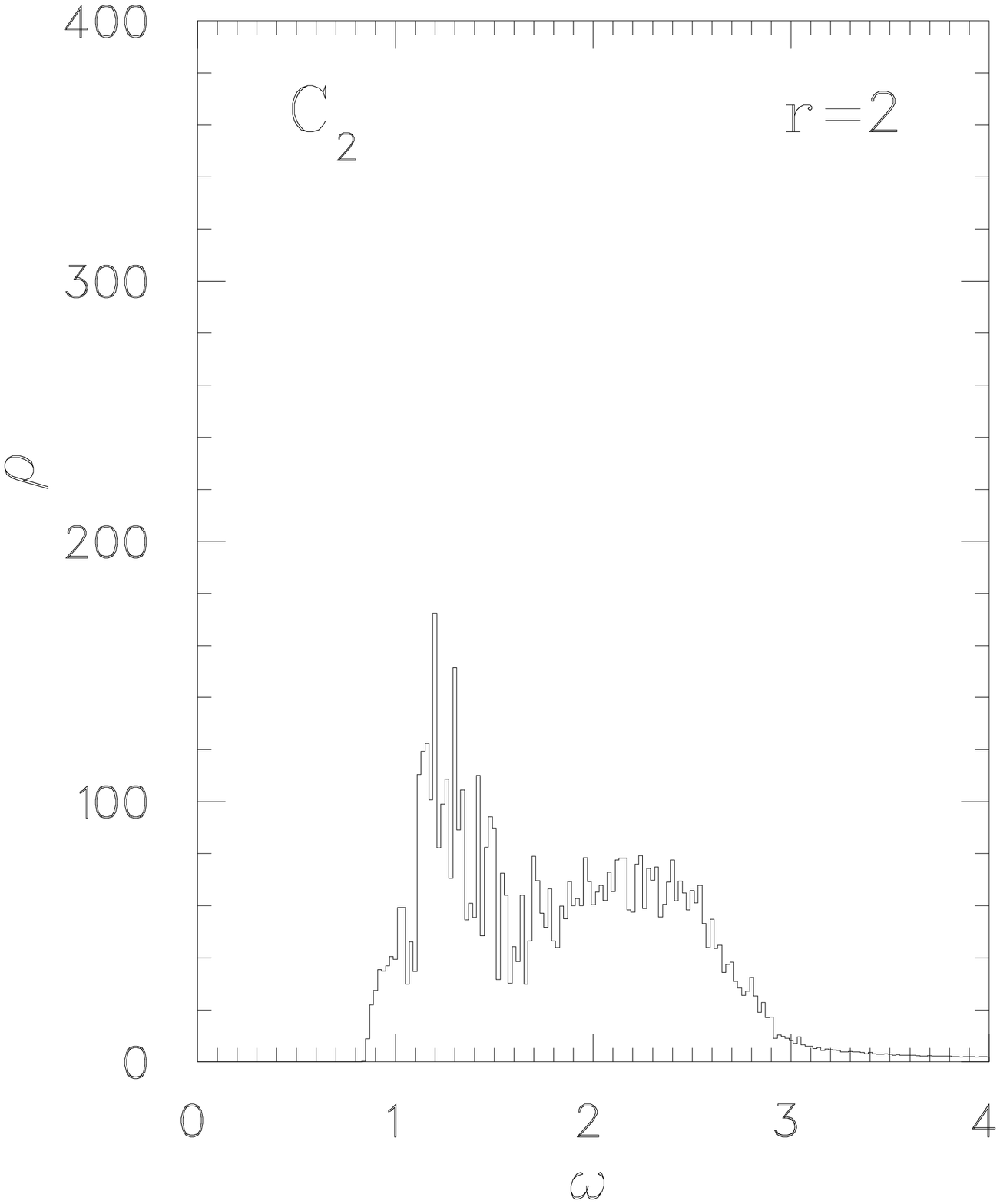}\vspace{2ex}\\
\includegraphics[angle=0,width=41.5mm]{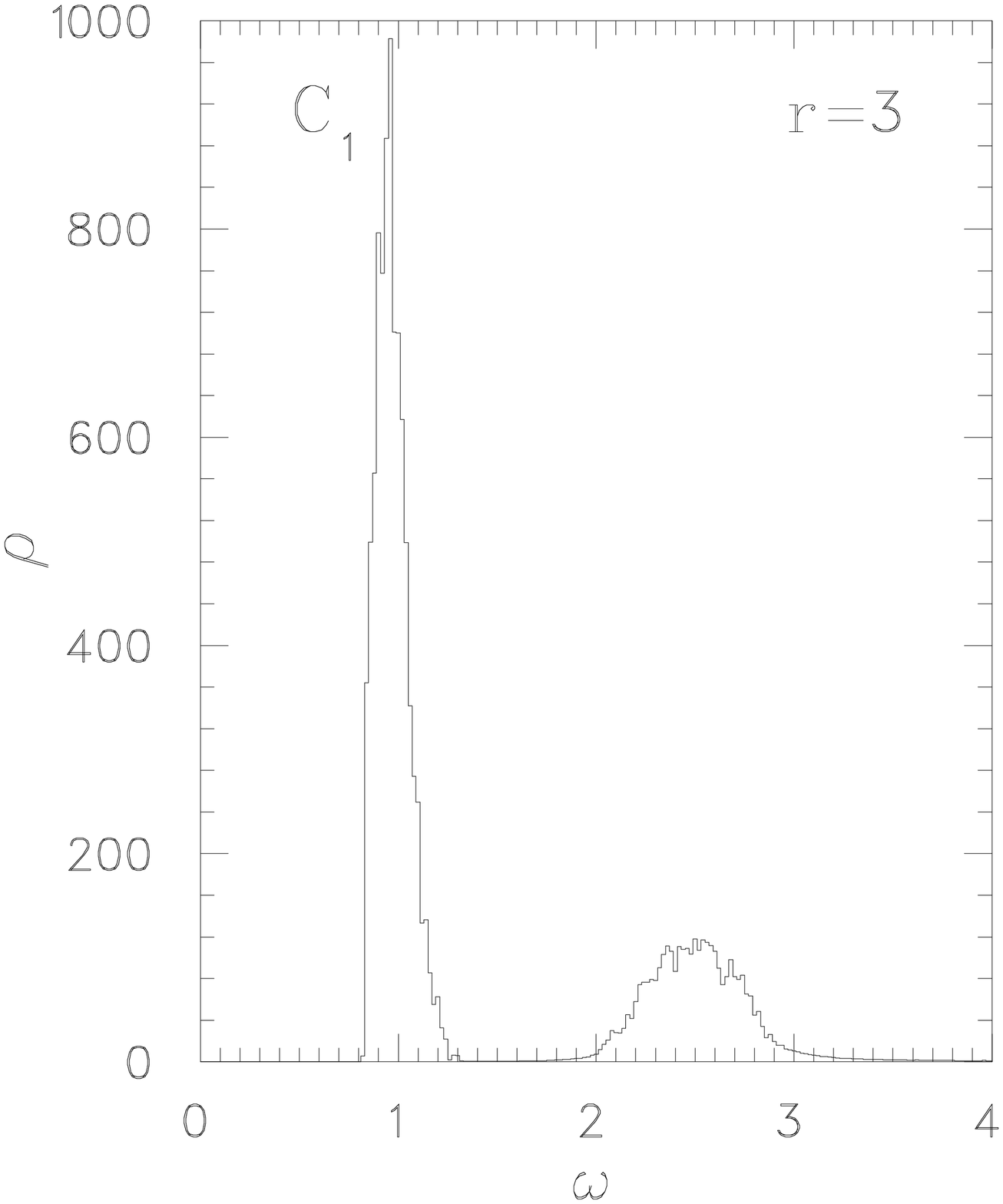}
\includegraphics[angle=0,width=41.5mm]{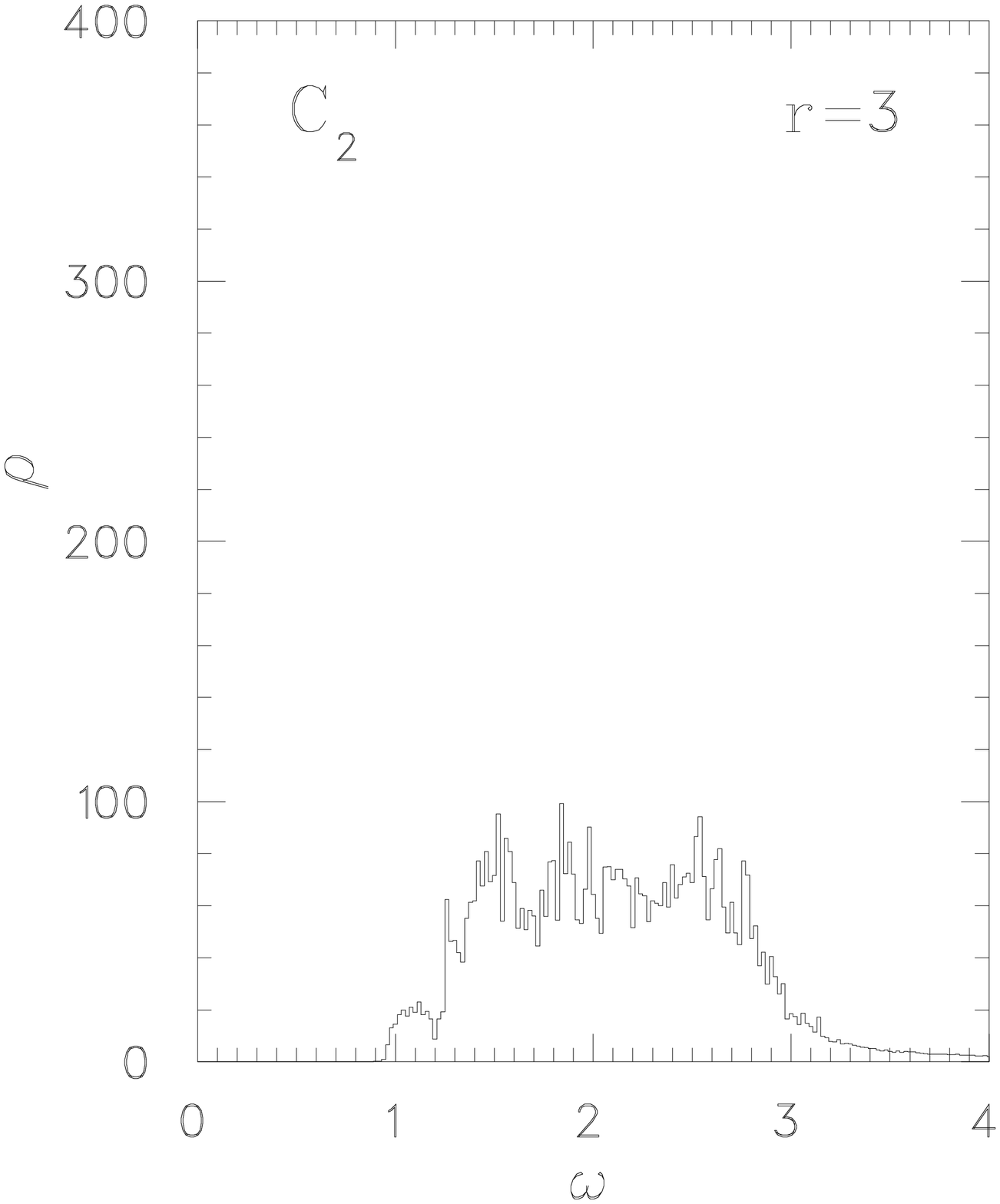}\vspace{2ex}\\
\includegraphics[angle=0,width=41.5mm]{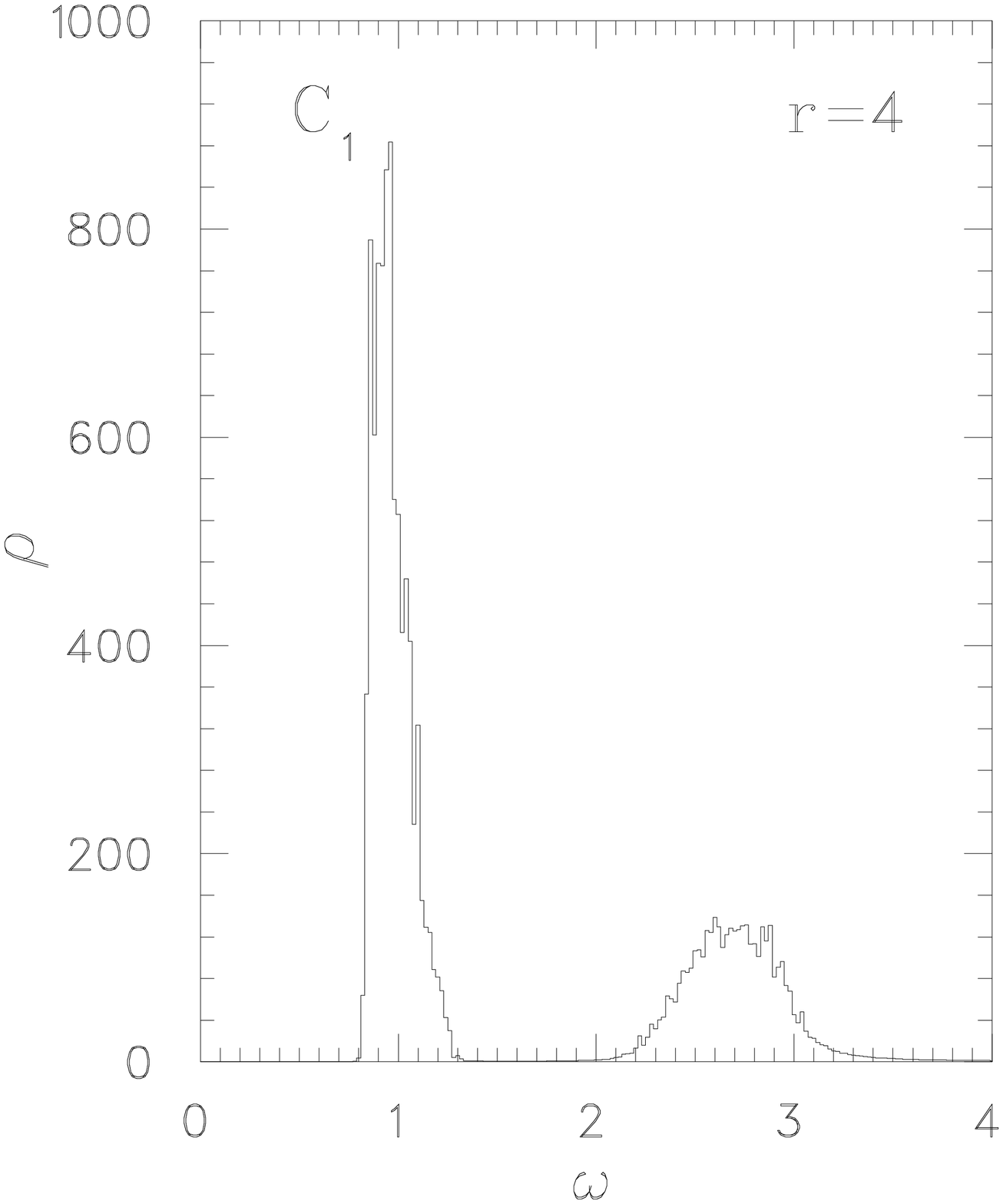}
\includegraphics[angle=0,width=41.5mm]{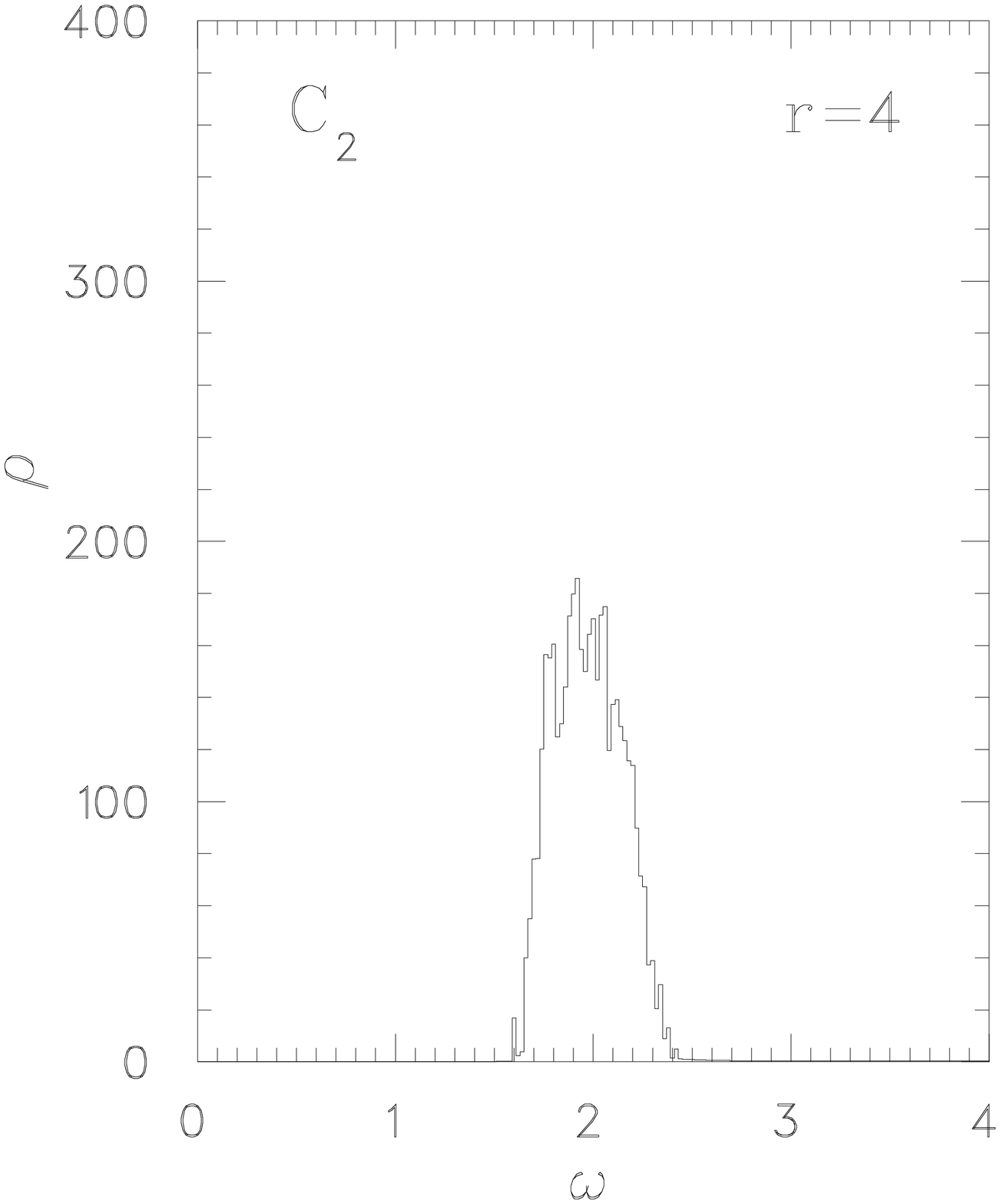}
\caption{\label{fig3}Spectral density functions $\rho$ of the eigenvalue
correlator (\protect\ref{vCv}) obtained by way of simulated annealing.
The graphs are the average over eight random annealing starts.
Spectra are shown side-by-side for the ground state correlator $C_1$ and
the excited state correlator $C_2$ for meson-meson
relative distances $r=1,2,3,4$.}
\end{figure}

\begin{figure}
\includegraphics[angle=0,width=41.5mm]{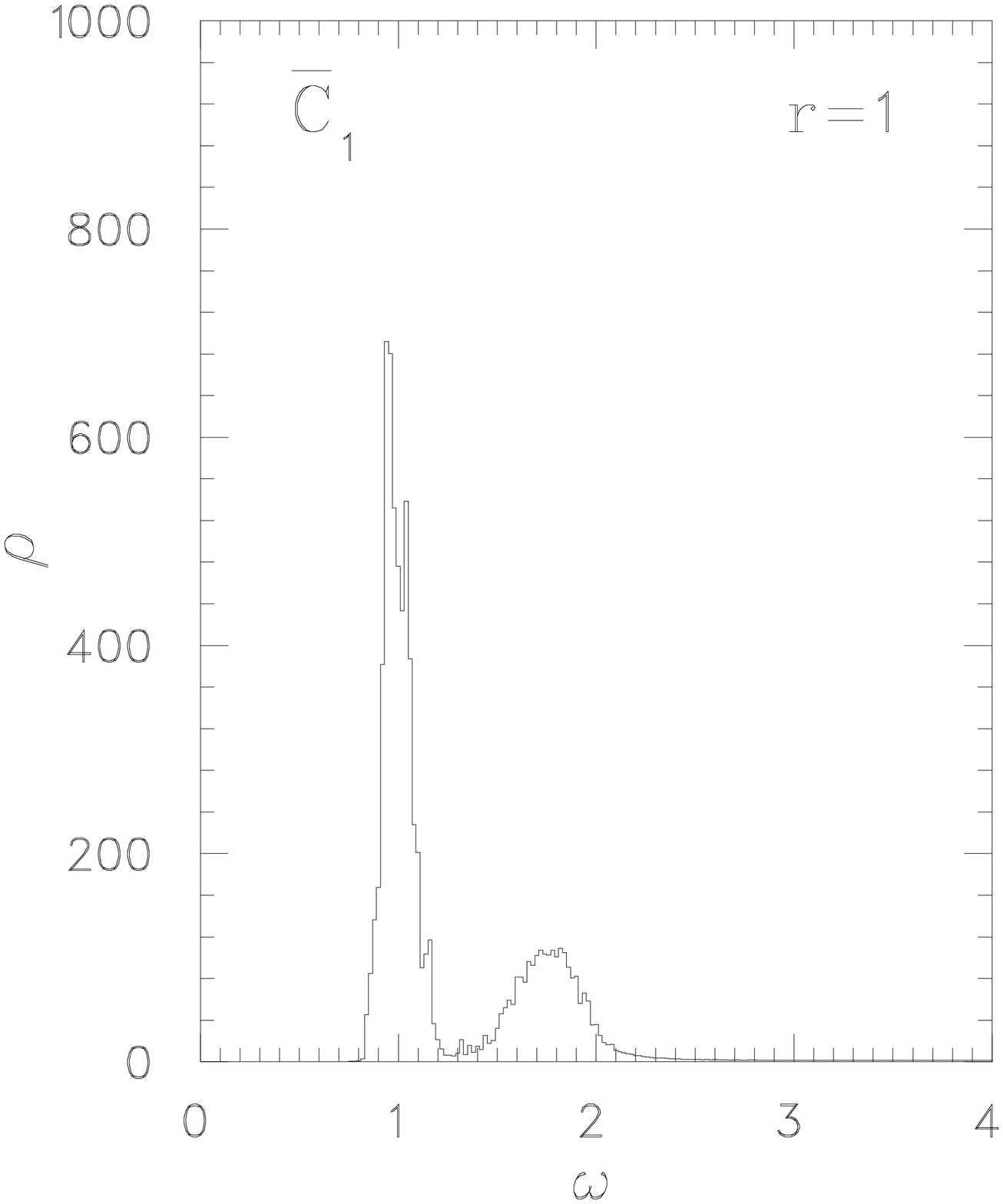}
\includegraphics[angle=0,width=41.5mm]{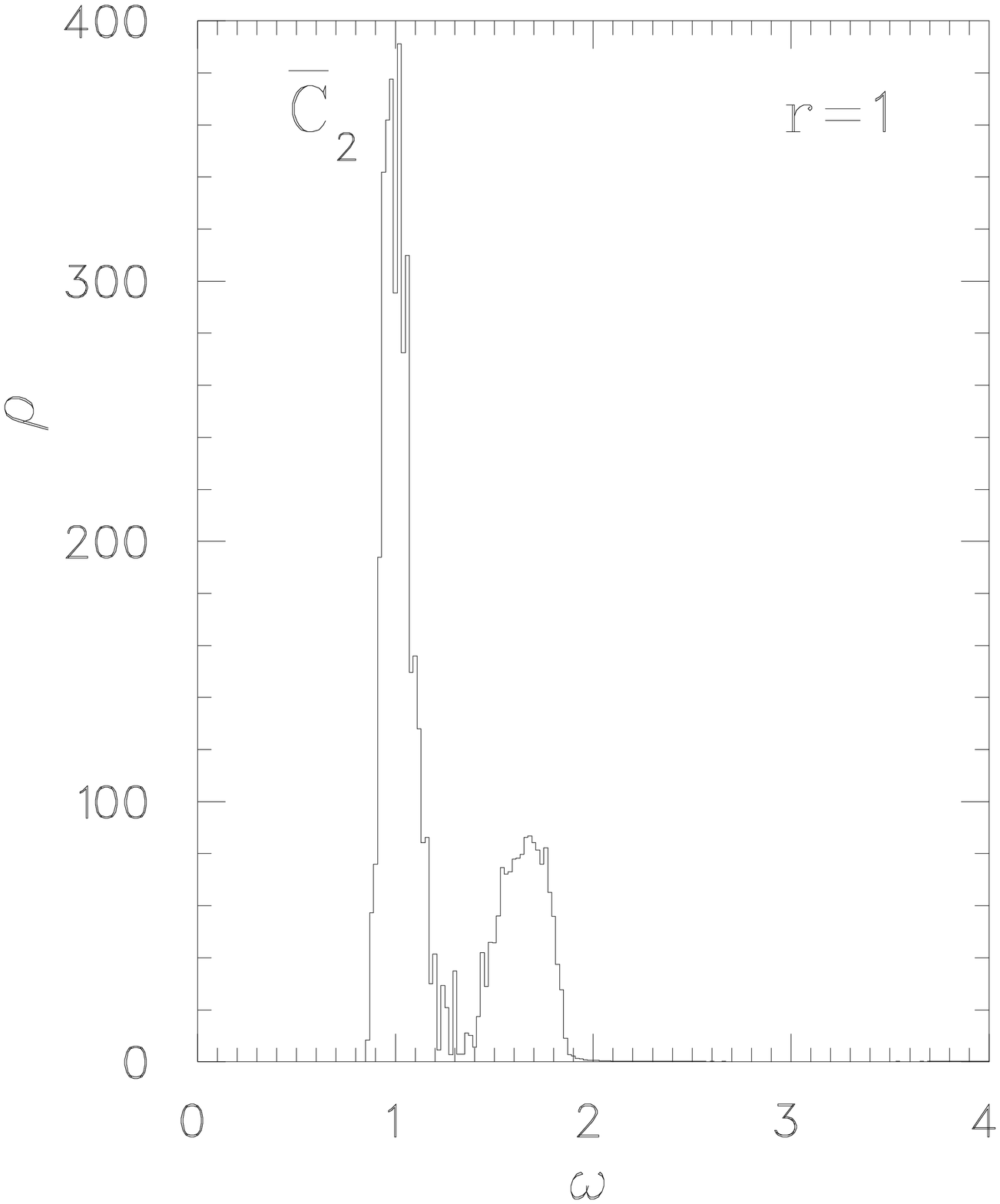}\vspace{2ex}\\
\includegraphics[angle=0,width=41.5mm]{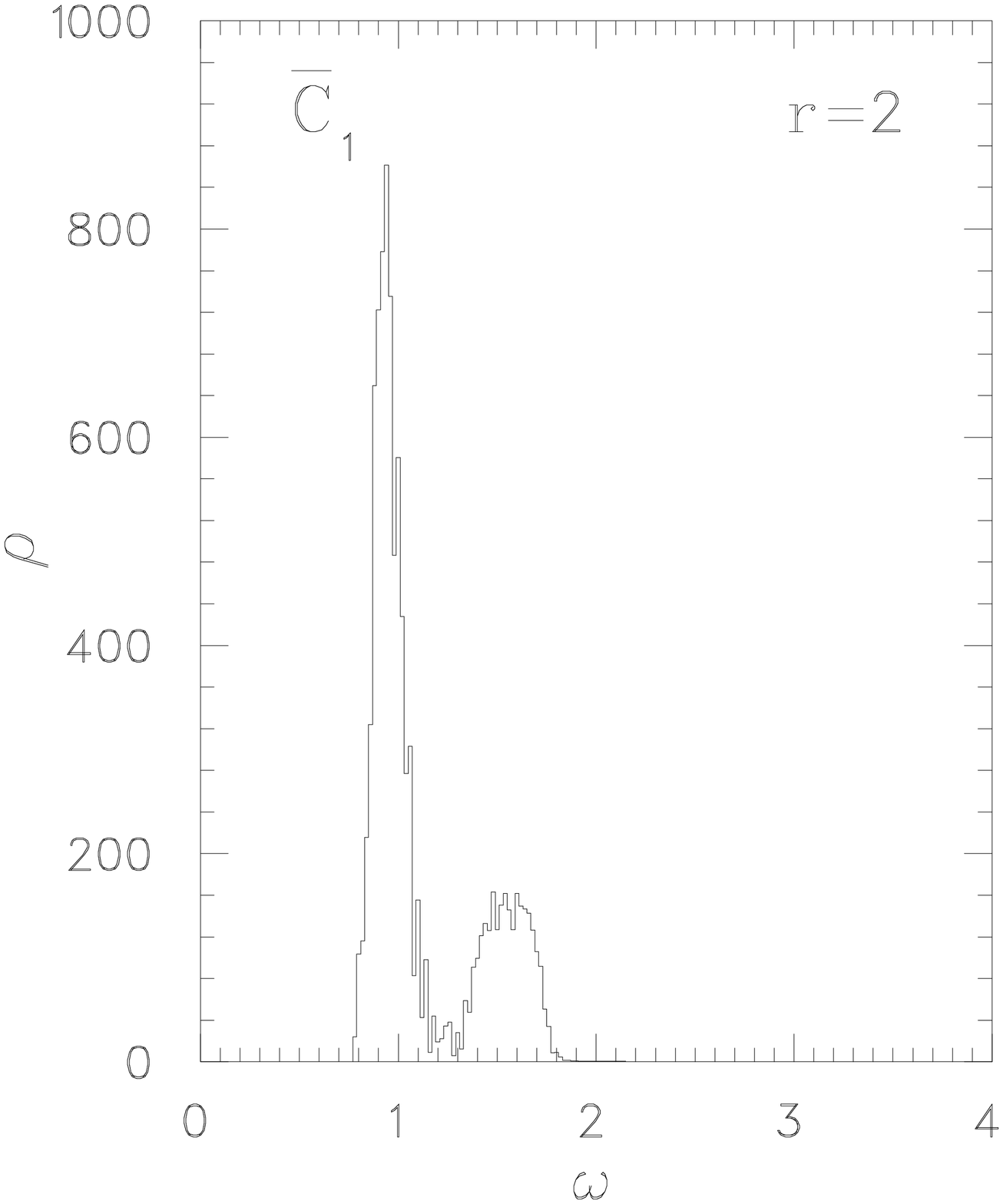}
\includegraphics[angle=0,width=41.5mm]{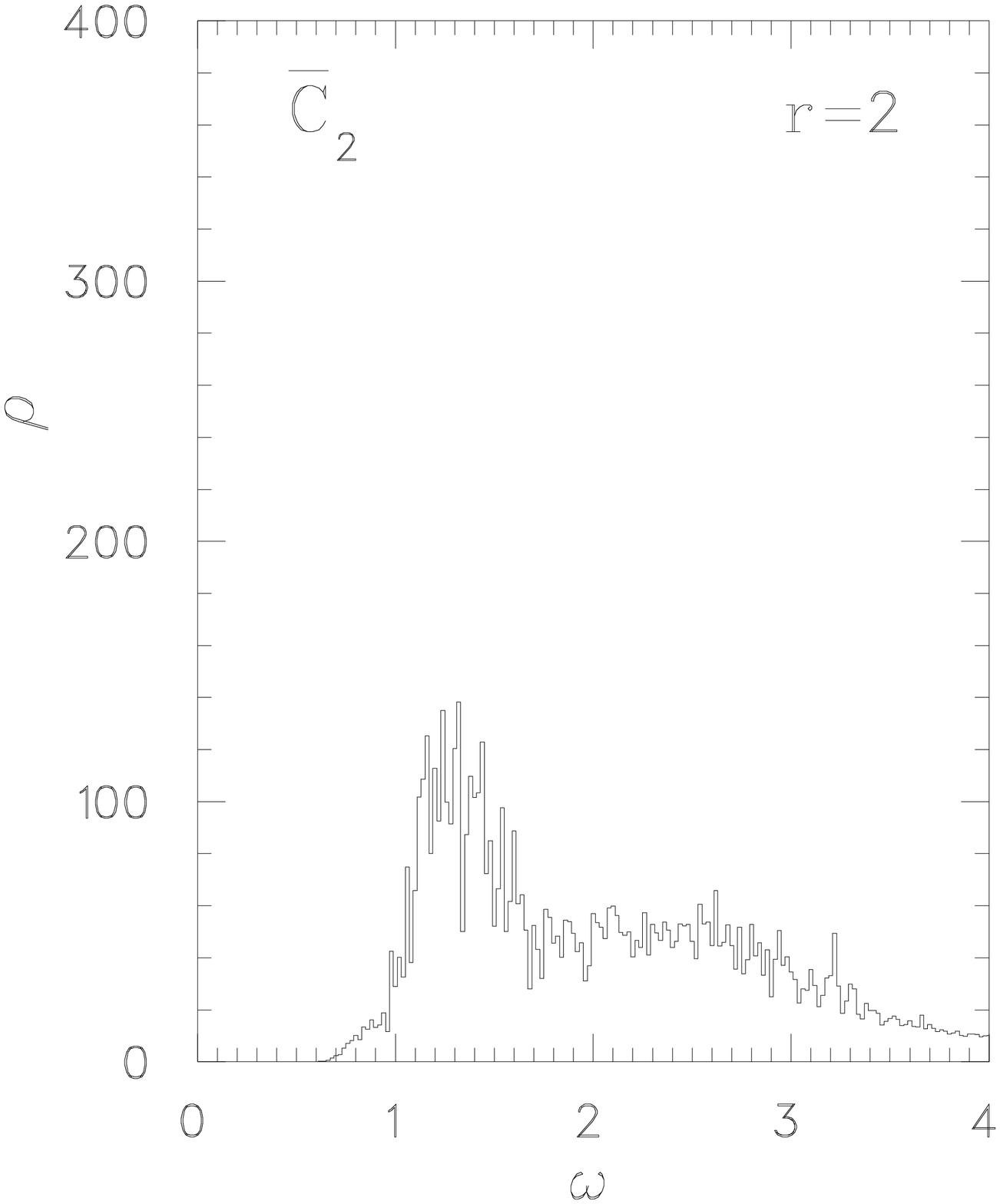}\vspace{2ex}\\
\includegraphics[angle=0,width=41.5mm]{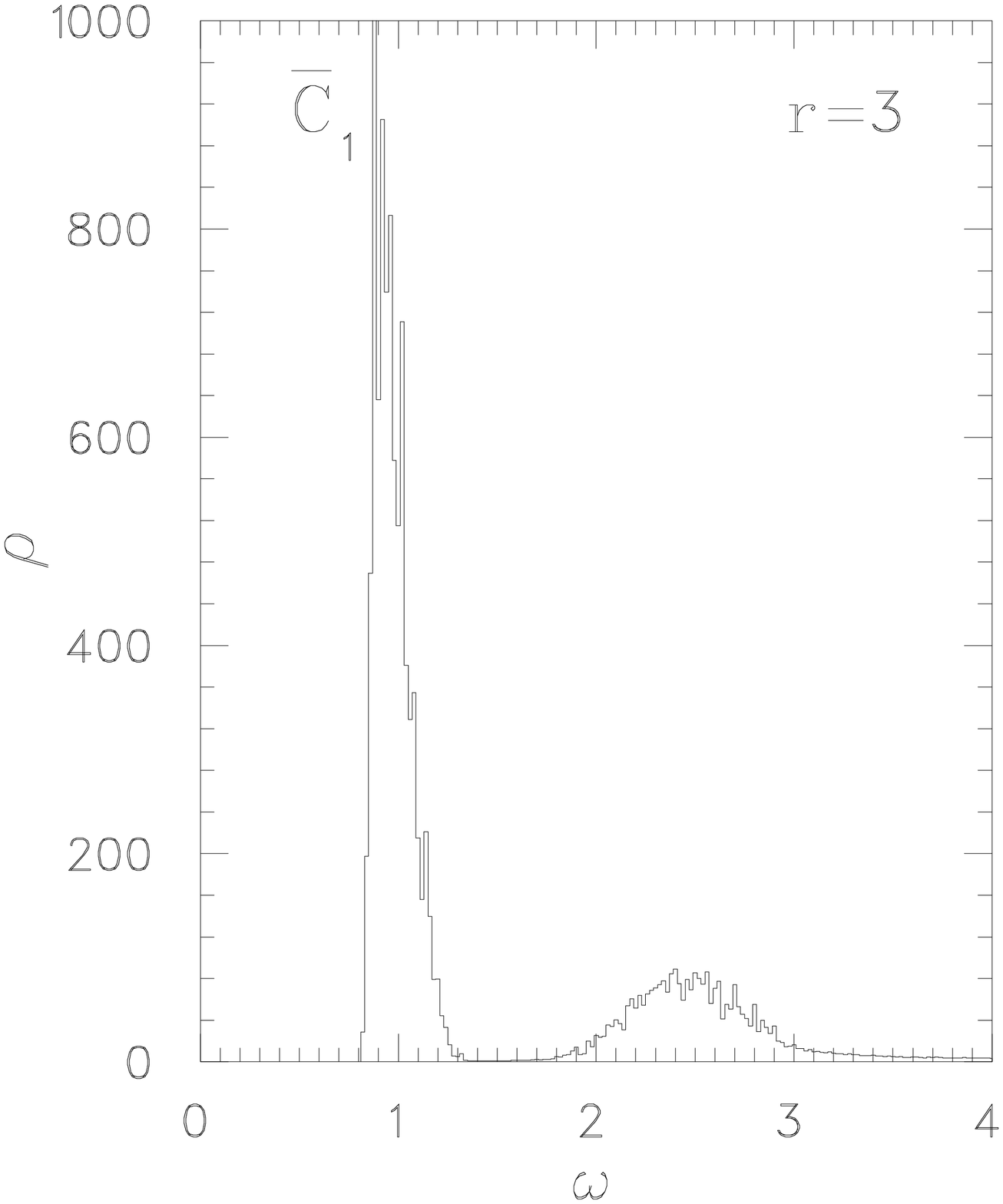}
\includegraphics[angle=0,width=41.5mm]{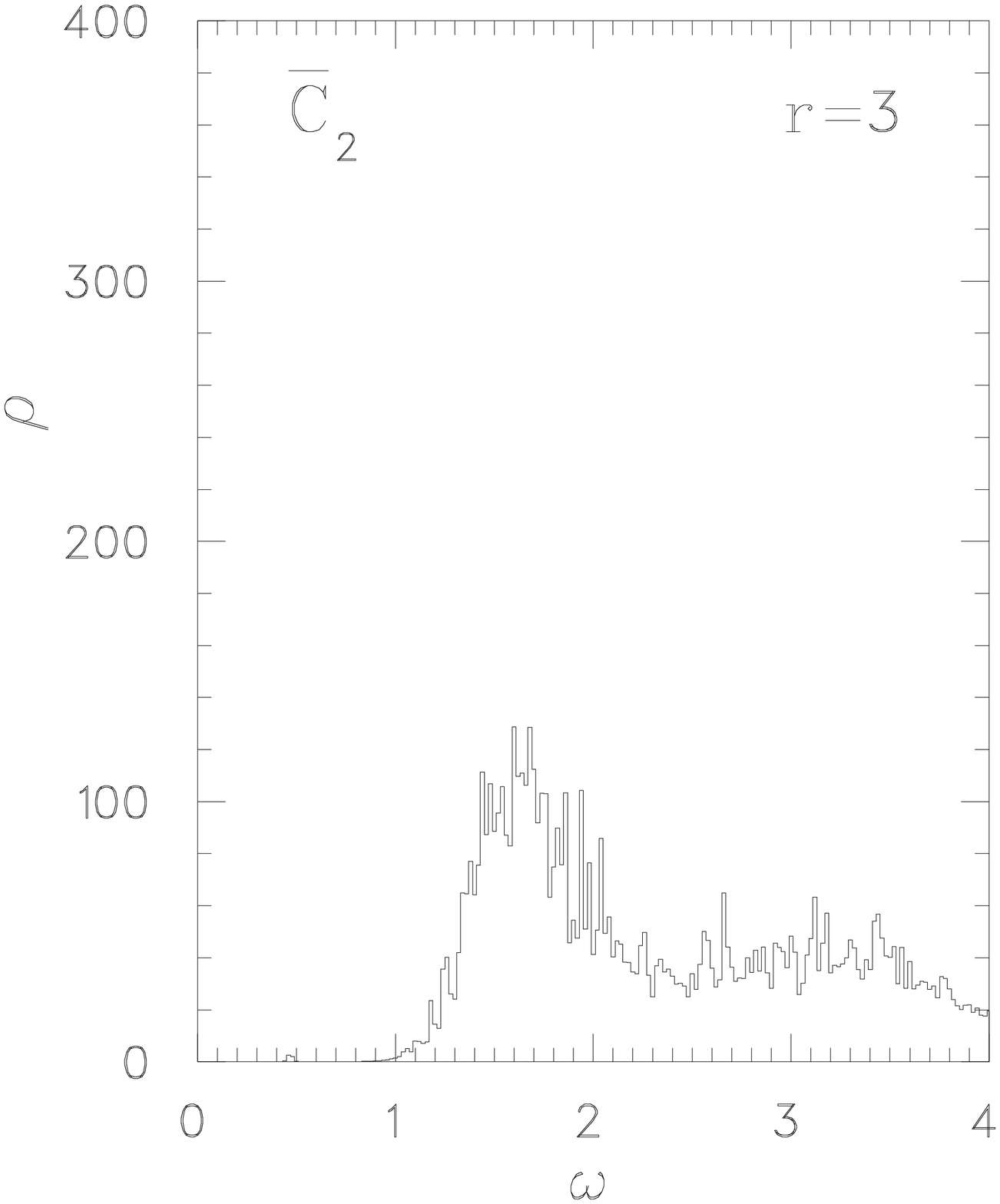}\vspace{2ex}\\
\includegraphics[angle=0,width=41.5mm]{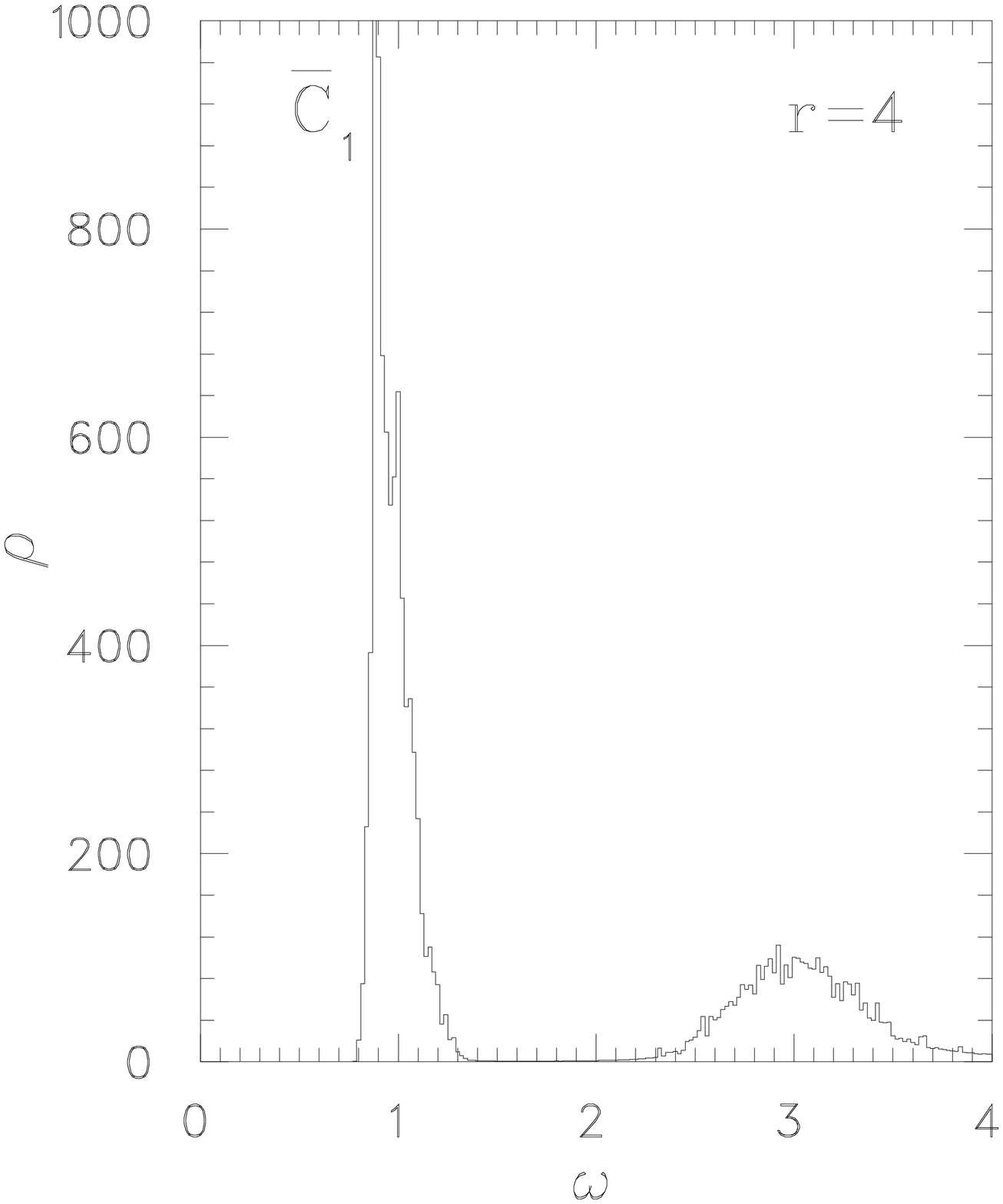}
\includegraphics[angle=0,width=41.5mm]{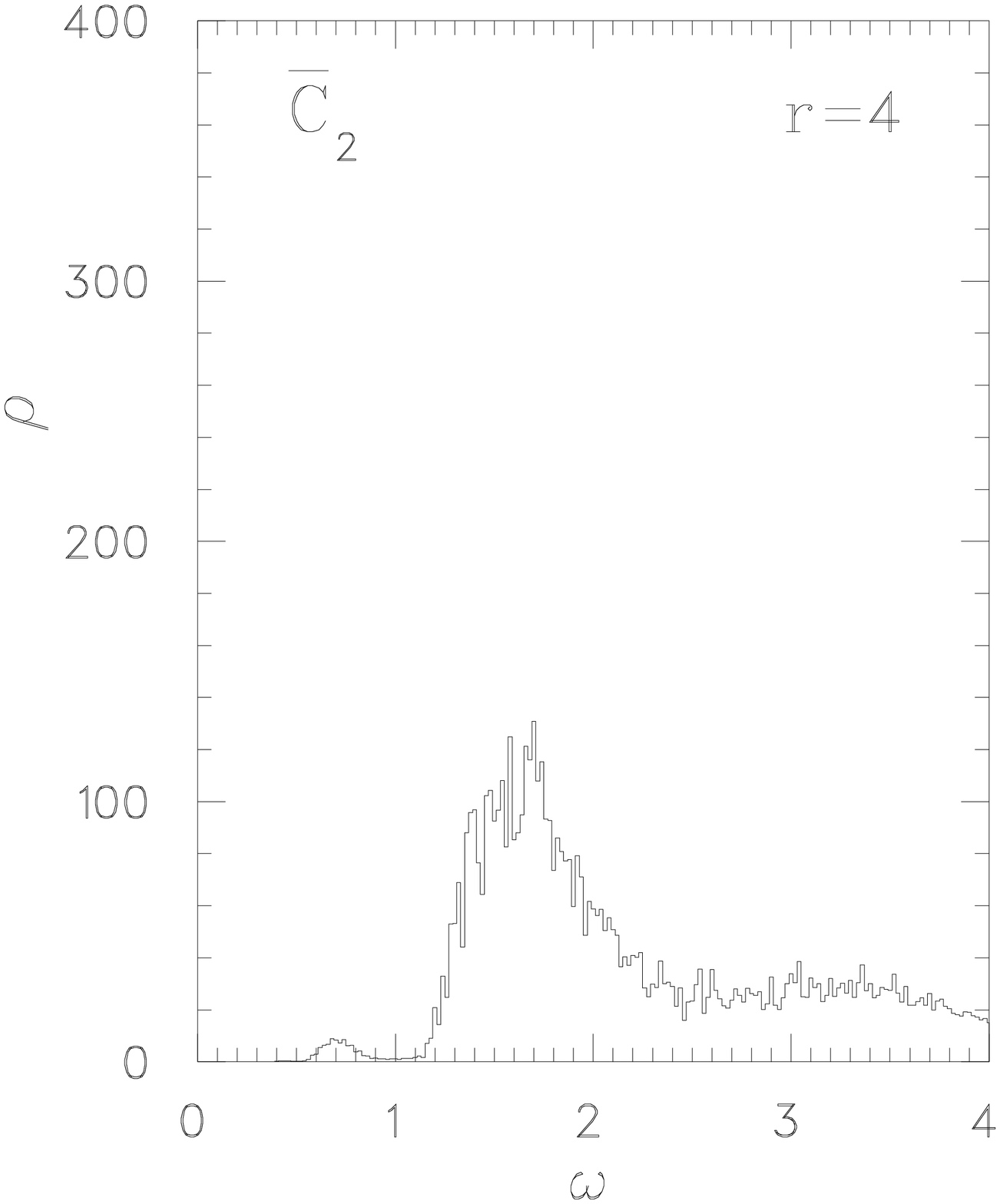}
\caption{\label{fig4}Spectral density functions as in Fig.~\protect\ref{fig3},
but for the asymptotic stabilized correlator functions (\protect\ref{bCb}).}
\end{figure}

Results of the MEM analysis are presented in Fig.~\ref{fig3} for the
eigenvalue correlators (\ref{vCv}) and in Fig.~\ref{fig4} for the asymptotic
stabilized correlators (\ref{bCb}).
As discussed in \cite{Fiebig:2002sp} the parameter choices are not critical.
Specifically, the entropy
strength is $\alpha=5.0\times 10^{-7}$, and the default model is constant with 
$m_k=10^{-12}, K_-\leq k\leq K_+$. The annealing schedule is given
in \cite{Fiebig:2002sp}. All spectral densities are averages over eight random
annealing starts. The $\omega$ discretization is set
by $\Delta\omega=0.02$, and $K_-=-100$, $K_+=+200$. 
Those numbers reflect the lattice design, like the energy cutoff ${a_t}^{-1}$,
and other considerations, see \cite{Fiebig:2002sp}.
The $\omega$ interval is larger and the resolution much finer compared to the
discretization used in \cite{Fiebig:2002sp}. Note that
the entire spectral mass range $-2.0\leq\omega\leq+4.0$, including backward
going propagation, is utilized in the spectral analysis. This is true of both
the eigenvalue correlators and the asymptotic stabilized correlators.
Most of the spectral structure, however,
is invisible on the linear scales used in Figs.~\ref{fig3} and \ref{fig4}, 
particularly for $\omega\leq 0$ (backward propagation).

As discussed in Sect.~\ref{sec:analysis}, in the limit $t\rightarrow\infty$ only
the lowest $\omega$ peak from each correlator $C_m(t,t_0), m=1\dots M$,
should be used to extract physical information.
We will refer to those as primary peaks.
Suppose primary peaks are seen in the spectral densities belonging to
$C_n(t,t_0)$, for $n=1\ldots N\leq M$.
We loosely characterize those by
\mbox{$\delta_n=\{\omega:\omega\in {\rm peak\ \#n}\}$}.
In \cite{Fiebig:2002sp} it was argued that low $\omega$ moments of the spectral
densities $\rho(\omega)$ can be reliably extracted. Specifically, these are
the peak volume $Z_n$, the mean energy $E_n$, and the width $\Delta_n$ of the peak 
\begin{eqnarray}
Z_n&=&\int_{\delta_n}d\omega\/\rho(\omega) \label{Zn}\\
E_n&=&Z_n^{-1}\int_{\delta_n}d\omega\/\rho(\omega)\omega \label{En}\\
\Delta_n^2&=&Z_n^{-1}\int_{\delta_n}d\omega\/\rho(\omega)
\left(\omega-E_n\right)^2\,. \label{Dn}
\end{eqnarray}
Note that the theorem (\ref{Lemma}) establishes the peak volume (\ref{Zn}) as
identical with the factors $Z_n$ introduced in Sect.~\ref{sec:analysis},
the caveat being that numerically extracted peaks have finite widths.
In particular, the peak volumes (\ref{Zn}) have the interpretation given by
(\ref{Zn2}), or by (\ref{Zn2a}) as optimal excitation probabilities.

In Figs.~\ref{fig3} and \ref{fig4} the (lowest mass) primary peaks
clearly dominate both ground state spectral density functions, $C_1$ and
$\bar{C}_1$, for each $r=1\ldots 4$.
With reference to (\ref{Zn2}) the secondary peaks at larger mass may
indicate that $|\!|\Pi\phi_1|\!|^2>0$, their volumes are smaller though.
It is much harder to make out a distinct peak structure in the spectral density
functions for the excited state eigenvalue correlators $C_2$.
We attribute this fact to strong statistical fluctuations of the eigenvector
components $v_{mi}(t,t_0)$, spoiling the signal. The asymptotic stabilized
excited state correlators $\bar{C}_2$ alleviate that problem, see Fig.~\ref{fig4}.
The spectral peaks of $\bar{C}_2$ are broad, for $r>1$.
This is an indication that the lattice data support them with only a few
consecutive points in the time correlation function. In other words, there
is not enough {\em information} in the data for distinct narrow
peaks to develop against the entropy background.
Although the peaks are wider, lower mass primary peaks are clearly distinguishable
for $r=1\ldots 4$ in Fig.~\ref{fig4}. 
In Tabs.~\ref{tab:raw} and \ref{tab:bar} we list the volumes, energies, and widths of
all primary peaks of Figs.~\ref{fig3} and \ref{fig4}.
For the excited states the $C_2$ data do not always clearly define a primary peak.
The corresponding numbers in Tab.~\ref{tab:raw} use $\omega$ cuts of
1.32, 1.58, 3.20 and 2.80, for $r=$ 1,2,3 and 4, respectively.
On the other hand, the $\bar{C}_2$ data originating with the asymptotic stabilized
correlators provide us with a much improved picture. There, by inspection
of Fig.~\ref{fig4}, the $\omega$ cuts are
1.28, 1.72, 2.20 and 2.46, for $r=1\ldots 4$, respectively.
Most of the subsequent discussion therefore uses results of the $\bar{C}$ analysis.

With an appropriate annealing schedule cooling fluctuations
at the final `temperature' can be made negligible.
The size of the statistical error from $N_U=708$ gauge configurations is
comparable to uncertainties from the MEM analysis. This was looked at
in \cite{Fiebig:2002sp}. Those uncertainties are introduced through using different
start configurations $\rho$ in the annealing process. 
There is no magic way to eliminate those, nor would it be desirable, because 
they test a property of the data set. Although theoretically $W[\rho]$ has a
unique absolute minimum sole knowledge of its location, say $\rho_{\min}$, is
deficient. It should be supplemented by having some notion about the
shallowness, or as the case may be, the distribution of local minima close
to $\rho_{\min}$ with values $W[\rho]$ not much different
from $W[\rho_{\min}]$ which the annealing algorithm may settle into.
Those manifest themselves in micro fluctuations (on a scale
of $\Delta\omega$) of the spectral density functions. Local minima close to
$\rho_{\min}$ appear to be numerous.
The numbers in Tabs.~\ref{tab:raw} and \ref{tab:bar} are from averages over eight
annealing start configurations, and the uncertainties are the corresponding
standard deviations. They are indicative of the spread of local minima of $W[\rho]$
in the vicinity of $\rho_{\min}$ and, ultimately of the uncertainty of the results.
\begin{table}
\caption{\label{tab:raw}Low $\omega$ moment features of the primary spectral peaks
extracted from the eigenvalue correlators $C_m(t,t_0), m=1,2$ for different relative
meson-meson distances $r$. The corresponding spectral density functions
are displayed in Fig.~\protect\ref{fig3}.
Listed are the peak volume $Z_m$, the peak energy $E_m$, and the peak width
$\Delta_m$, for $m=1,2$, as defined in (\protect\ref{Zn})--(\protect\ref{Dn}).
All entries are averages over eight random annealing start configurations, the
uncertainties are the corresponding standard deviations.}
\begin{ruledtabular}
\begin{tabular}{ccccccc}
$r$ & $Z_1$ & $E_1$ & $\Delta_1$ & $Z_2$ & $E_2$ & $\Delta_2$ \\
\colrule
1.0 & 124.3(5) & 0.985(2) & 0.062(5) &  97.5(1)  & 1.277(2) & 0.263(3) \\
2.0 & 141.(2)  & 0.947(5) & 0.066(9) & 129.0(4)  & 1.851(6) & 0.595(9) \\
3.0 & 176.(2)  & 0.971(4) & 0.083(9) & 119.7(5)  & 2.109(8) & 0.569(7) \\
4.0 & 174.(1)  & 0.970(4) & 0.093(6) &  84.7(1)  & 1.983(2) & 0.203(3) \\
\end{tabular}
\end{ruledtabular}
\end{table}
\begin{table}
\caption{\label{tab:bar}Low $\omega$ moment features of the primary spectral peaks
like Tab.~\protect\ref{tab:raw}, but for the asymptotic stabilized correlators
$\bar{C}_m(t,t_0), m=1,2$ of (\ref{bCb}). The corresponding spectral density
functions are displayed in Fig.~\protect\ref{fig4}.}
\begin{ruledtabular}
\begin{tabular}{ccccccc}
$r$ & $\bar{Z}_1$ & $\bar{E}_1$ & $\bar{\Delta}_1$ &
      $\bar{Z}_2$ & $\bar{E}_2$ & $\bar{\Delta}_2$ \\
\colrule
1.0 & 107.(1)  & 0.997(3) & 0.072(5) & 67.2(9) & 1.002(5) & 0.072(9) \\
2.0 & 141.(2)  & 0.951(4) & 0.074(7) & 44.(1)  & 1.29(1)  & 0.191(8) \\
3.0 & 175.7(9) & 0.974(3) & 0.090(7) & 60.(2)  & 1.63(1)  & 0.246(9) \\
4.0 & 162.6(6) & 0.969(2) & 0.094(3) & 65.(1)  & 1.64(1)  & 0.263(9) \\
\end{tabular}
\end{ruledtabular}
\end{table}

In view of the broader peaks in Figs.~\ref{fig3} and \ref{fig4} it may be argued
that the widths $\Delta_2$ and $\bar{\Delta}_2$ supersede the standard deviations
derived from annealing start configurations as a useful indicator for the
uncertainties of the average energies $E_2$ and $\bar{E}_2$, see Tabs.~\ref{tab:raw}
and \ref{tab:bar}. In fact there are at least three types of indicators:
(i) the gauge field configuration statistical error,
(ii) the annealing start configuration standard deviation, and
(iii) the peak width.
They all point at different aspects of the uncertainty of the $E_n$.
For example, the $\Delta_n$ convey the aspect of {\em information content} of
the data in the sense of \cite{Sha49}.
A deeper discussion is beyond the scope of this work.
For the current results the gauge field configuration statistical error is
typically smaller than 2\% and thus much less than the other two measures of
uncertainty. Then, the annealing start configuration standard deviation in most
cases is much less than the peak width. We here adopt the point of
view that, since Bayesian inference is built on the {\em information content} of
the data, the peak width is the appropriate measure of uncertainty.
However, for reference to (i) refer to \cite{Fiebig:2002sp}, and we include
(ii) in the results below as appropriate.

\begin{figure}
\includegraphics[angle=90,width=82mm]{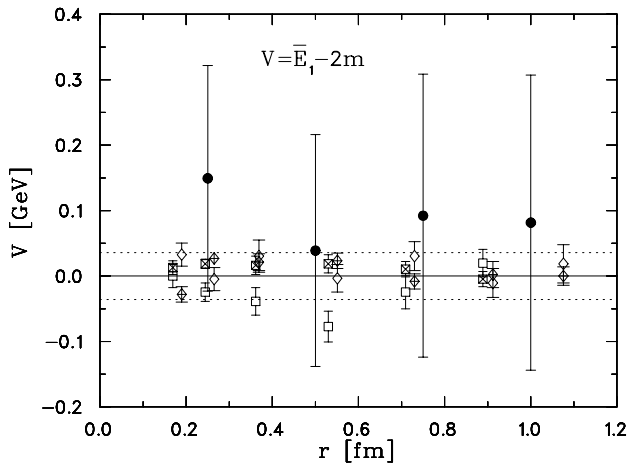}
\caption{\label{fig5}Ground state energies of the meson-meson system for various
relative distances $r$.
The mass $2m$ of the non-interacting system and the
gauge configuration statistical error, both from \protect\cite{Fiebig:2002sp}, are
shown as horizontal lines.
The solid plot symbols correspond to the energies $\bar{E}_1$ of this work,
the uncertainties are spectral peak widths $\bar{\Delta}_1$.
For comparison we show the $I=1$ and $S=0,1$ data for two lattices
compiled from \protect\cite{Michael:1999nq}.}
\end{figure}

The ground state energies $\bar{E}_1$ of the meson-meson system are plotted in
Fig.~\ref{fig5} together with the mass $2m$ and the error band for
non-interacting mesons. The mass $m=0.4676$ and the gauge configuration statistical
error $\Delta m=0.0075$ are those from $m_{\rm eff,0}$ in \cite{Fiebig:2002sp}.
Error bars on the data points are the peak widths $\bar{\Delta}_1$.
Although the errors extend well below $V=0$, a systematic tendency for 
repulsion, averaging about $\approx 90{\rm MeV}$ above $2m$, is apparent.
These features are consistent with previous lattice studies of the pseudoscalar
meson-meson system in the same isospin channel \cite{Fiebig:1999hs}.

We also display in Fig.~\ref{fig5} results from \cite{Michael:1999nq} which relate
to the current discussion. In \cite{Michael:1999nq} operators are characterized by
the isospin and spin symmetry of the light quarks. Because in the present work all
the light quarks have the same flavor, only the symmetric isospin combination is
relevant; in the notation of \cite{Michael:1999nq} this is $I=1$.
Figure~\ref{fig5} shows their $S=0$ and $S=1$ results from two lattices.
The uncertainties are gauge configuration statistical errors and are thus
much smaller than our peak width error bars. The energy values populate
a $\pm 50{\rm MeV}$ band around $V=0$.
It should be noted that the authors of \cite{Michael:1999nq} go through extraordinary
length to obtain the `best' ground state energies possible by means of diagonalizing
a matrix of operators (similar to $\Phi_1$) with different fuzzing levels of the gauge
fields. This goes beyond the widely used practice to employ a few iterative steps,
say $N$, of gauge link fuzzing \cite{Alb87a} and quark field
smearing \cite{Alexandrou:1994ti}. A reasonable choice for the number of iterations
is to make $Na_s$ about the radius of the hadron considered.
(In our case, with $a_s=0.25{\rm fm}$, this would be $N=4$, for example.) 
The result is a spatially extended operator, spreading across a hadron volume,
that makes correlation functions assume asymptotic behavior at `earlier' time slices.
This is the procedure adopted in the current work.
This practice also has the side effect of lowering the
ground state energy obtained from the simulation because, numerically,
contaminations from excited states are reduced. This effect is of course enhanced
if the ground state is extracted from diagonalizing a correlation matrix from
operators of different fuzzing levels, as employed in \cite{Michael:1999nq}.
The size of the effect can be seen in Fig.~\ref{fig5}.
We note that the physics goal of the authors of \cite{Michael:1999nq} is
to investigate if the $\cal BB$ system can possibly be bound.
In that context even small effects on the ground state energy are vital,
so the elaborate matrix fuzzing procedure is justified.
However, the physics goals of
this work are completely different. Studying the interaction mechanism rests on
a comparison of ground and excited state energies, as $r$ changes.
Because the excited states energy levels are substantially larger, tiny
shifts in the ground state energies are irrelevant.
  
\begin{figure}
\includegraphics[angle=90,width=82mm]{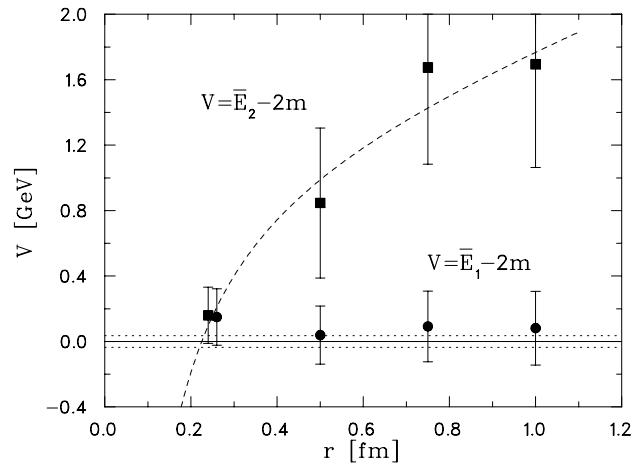}
\caption{\label{fig9}
Energies of the excited (squares) and ground (circles) meson-meson states
relative to twice the single meson mass in physical units. The uncertainties
are peak widths from the spectral analysis. The data points at $r=0.25{\rm fm}$
are shifted sideways slightly, for better visibility.
The curve is a fit with $y=-a/r+br+c$, see text.}
\end{figure}

This situation is reflected in Fig.~\ref{fig9} where we show excited state
and ground state energies together, versus the relative meson-meson distance. 
The energy $\bar{E}_2$ drops considerably as $r$ decreases.
At large $r$, according to Tab.~\ref{tab:sfit} and Fig.~\ref{fig11R},
the operator $\Phi_2(t)$ is mainly responsible for the values of $\bar{E}_2$.
Bearing in mind that $\Phi_2(t)$, see (\ref{Phi2}) and Fig.~\ref{fig1},
involves color charges separated by a distance $r$, which should be confined,
we have fitted the data with the model $y=-a/r+br+c$, where the parameter
$b$ was fixed to the string tension, $\sqrt{b}=0.44{\rm GeV}$
or $b=0.968{\rm GeV}/{\rm fm}$. The remaining fit returns
$a=0.30(14)\,{\rm GeV\,fm}$, $c=1.09(52)\,{\rm GeV}$, where the uncertainties
are variances. The resulting curve is shown in Fig.~\ref{fig9}.
Around $r\approx 0.2{\rm fm}$ a level crossing between the ground
and excited state energies of the meson-meson system apparently occurs.
From the viewpoint of an adiabatic potential this distance defines the
transition point (avoided level crossing) between a weakly repulsive
interaction mostly mediated by quark exchange, and strongly attractive
interaction from gluon exchange degrees of freedom. The transition distance
is consistent with Fig.~\ref{fig11R}, where at a point somewhat smaller than
$r\approx 0.5$ marks equal $s$ values for the $\Phi_1$ and $\Phi_2$ components
of the asymptotic eigenvectors $v_1$ and $v_2$. 
The picture emerging from the current results thus is that of heavy-light
pseudoscalar mesons having a radius of about $R\approx 0.1{\rm fm}$,
with a sharp boundary, as probed by their mutual interaction.
Note that the dramatic change of the adiabatic potential, at the distance of
the avoided level crossing nevertheless comes from a smooth transition
between the interaction mechanisms, as manifest in Fig.~\ref{fig11R},

By extrapolation it appears that below $r\approx 0.2{\rm fm}$ the
interaction turns attractive.
This observation is in line with \cite{Michael:1999nq} where
strong attraction at $r=0$ due to gluonic effects is observed, and
also with results in \cite{Mihaly:1997ue,Fiebig:1999hs}.
A simple explanation is suggested by the $SU(3)$ color content of the
two-body operators. Using standard nomenclature \cite{Lic78,Clo79} we
note that only the singlet from
\begin{eqnarray}
\overline{\bf 3} \otimes \bf 3 &=& \bf 8 \oplus \bf 1
\label{bar3x3}\end{eqnarray}
is used in the construction of (\ref{Phi1}). Thus, generically, the
color-source structure of the operator $\Phi_1$ is
\begin{equation}
\Phi_1^{(\bf 1)}(1,2) \sim \phi^{(\bf 1)}(1) \, \phi^{(\bf 1)}(2) \,,
\label{Phi11}\end{equation}
where 1 and 2 denote the two color sources.
The decomposition of the product of two color octets
\begin{eqnarray}
\bf 8 \otimes \bf 8 &=& \bf 27 \oplus \bf 10 \oplus \bf 8 \oplus \bf 8
\oplus \overline{\bf 10} \oplus \bf 1
\label{8x8}\end{eqnarray}
also contains a singlet and therefore mixes with (\ref{Phi11}).
The construction of (\ref{Phi2}) involves two gauge link paths.
Thus in some sense the color-source structure of $\Phi_2$ is,
schematically, described by the Clebsch-Gordon series
\begin{equation}
\Phi_2^{(\bf 1)}(1,2) \sim \sum_{i,j}
\left( \begin{array}{c} {\bf 1} \\ \phantom{C} \end{array} \right.
\left| \begin{array}{cc} \bf 8 & \bf 8 \\ i & j \end{array} \right) 
\phi^{(\bf 8)}_i(1) \, \phi^{(\bf 8)}_j(2) \,.
\label{Phi88}\end{equation}
The interaction energies for one-gluon exchange in those states are proportional
to the expectation values of
$2\vec{F}(1)\cdot\vec{F}(2) = (\vec{F}(1)+\vec{F}(2))^2-\vec{F}(1)^2-\vec{F}(2)^2$
where $\vec{F}^2$ is an $SU(3)$ Casimir operator \cite{Clo79}.
A simple calculation gives
\begin{eqnarray}
\langle 2\vec{F}(1)\cdot\vec{F}(2)\rangle_{\Phi^{(\bf 1)}_1} &=&  0    \label{FF0}\\
\langle 2\vec{F}(1)\cdot\vec{F}(2)\rangle_{\Phi^{(\bf 1)}_2} &=& -6\,, \label{FF6}
\end{eqnarray}
which indicates (possibly strong) attractive interaction at small $r$
such as described by excitations due to $\Phi_2$.
This is only a schematic picture, but is it appears to be consistent with
the results of the lattice simulation.\footnote{Note that $r=0$ is a special
case since there is no distinction in the color structure of (\protect\ref{Phi1})
and (\protect\ref{Phi2}), in fact $\Phi_1=\Phi_2$ at that point.}
 
\begin{figure}
\includegraphics[angle=0,width=42mm]{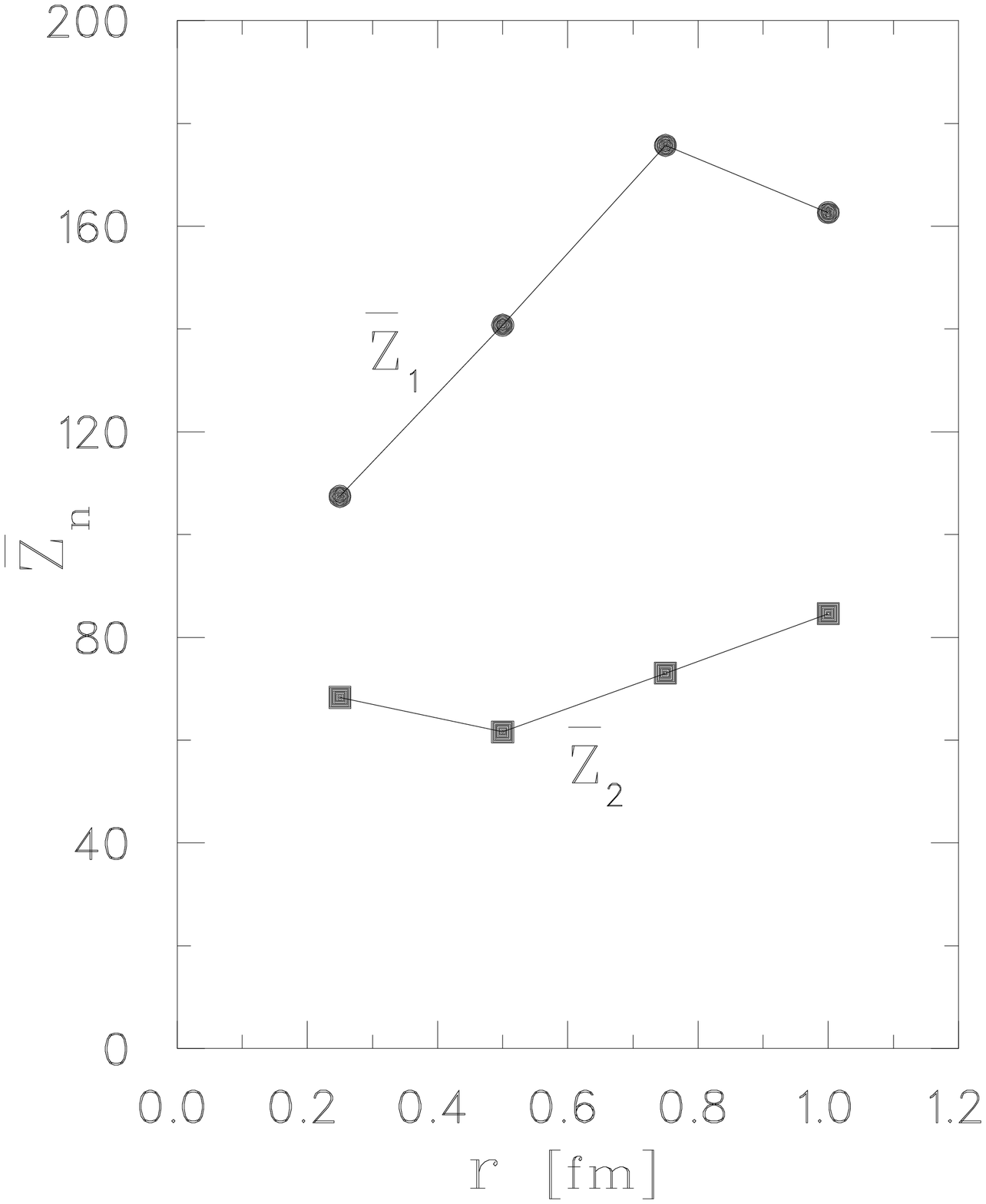}\vspace{0mm}
\includegraphics[angle=0,width=42mm]{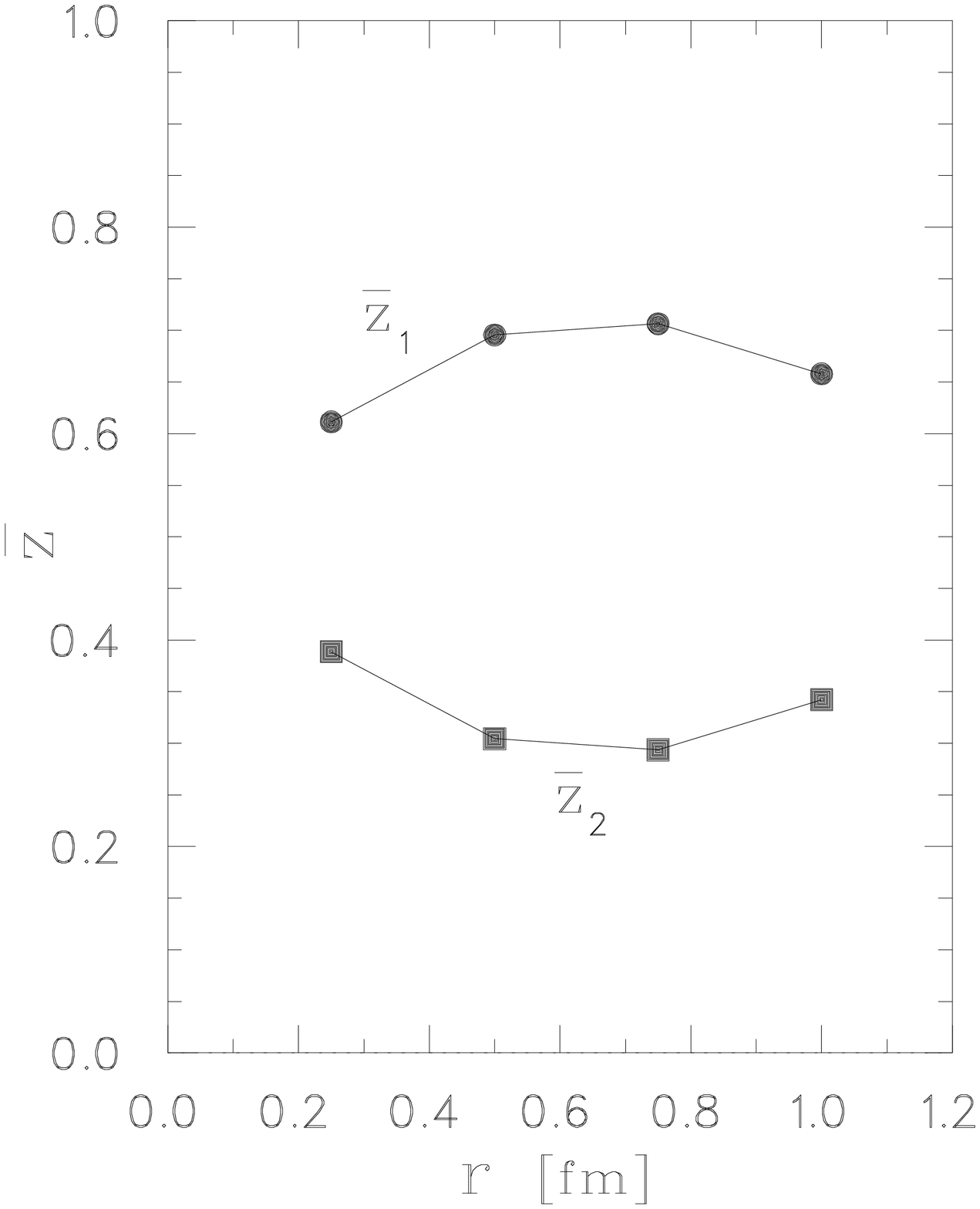}
\caption{\label{fig6}Dependence of the peak volumes $\bar{Z}_n$ and the
normalized peak volumes $\bar{z}_n$ for the ground and excited states $n=1,2$,
The uncertainties, which are
annealing start standard deviations, are small and obscured by the plotting symbols.}
\end{figure}

Finally, we turn to the transition matrix elements. Equations (\ref{Zn2}) and (\ref{Zn2a})
relate the peak volumes $Z_n$ to (unnormalized) excitation probabilities of the states
$|n\rangle$,
\begin{subequations}
\begin{eqnarray}
Z_n&=&\left|\langle n|\:v_{n1}\Phi_1(t_0)+v_{n2}\Phi_2(t_0)\:|0\rangle\right|^2\label{Zopa}\\
&\leq&\left|\langle n|\Phi_1(t_0)|0\rangle\right|^2+
\left|\langle n|\Phi_2(t_0)|0\rangle\right|^2\label{Zopb}\,.
\end{eqnarray}
\end{subequations}
The coefficients $v_{ni}$ in (\ref{Zopa}), being eigenvector components, ensure
that $Z_n$ is maximal within the linear space of the available operators
$\Phi_i(t_0), i=1,2$, see Sect.~\ref{sec:analysis}. 
Accordingly we interpret $Z_n$, or
\begin{equation}
z_n=Z_n/(Z_1+Z_2)\,,
\label{zn}\end{equation}
as measures for the effectiveness of the set of operators to excite the state $|n\rangle$. 
The corresponding results, compiled from Tab.~\protect\ref{tab:bar}, are displayed in
Fig.~\ref{fig6}.
The decrease of $\bar{Z}_1$ as $r$ becomes smaller means that the operators $\Phi_{1}$
and $\Phi_{2}$ start failing to capture some physics of the two-meson ground
state at small relative distance. This is a sign of an emerging interaction mechanism
that is is not well represented by $\Phi_{1,2}$. Other operators should eventually be
added to the set. From Fig.~\ref{fig6} we also see
that the operator set $\Phi_{1,2}$ is about 60--70\% effective in
creating the two-meson ground state, and the rest is left to excited state creation.
It is possible that gauge link fuzzing and quark field smearing is responsible for
the enhanced ground state presence. The normalized excitation rates are essentially
independent of $r$, as Fig.~\ref{fig6} shows. 

\section{\label{sec:conclusion}Summary and conclusion}

In an attempt to learn about the physics of hadronic interaction we have studied
the spectrum of a meson-meson system consisting of two light and two static quarks.
The static quarks define the relative distance $r$ between the mesons.
We have supplemented the commonly used local two-meson operator
$\Phi_1(t)$ with a nonlocal one $\Phi_2(t)$ which is built from spatially extended
color singlets. The eigenvalues of the corresponding $2\times 2$ time correlation matrix
were used in the spectral analysis thus allowing operator mixing.
We have employed the maximum entropy method (MEM), a form of Bayesian
inference, to extract the spectral densities of the ground and the excited states from
the time correlation matrix.

The spectral analysis yields both energies of the ground and excited states of the
meson-meson system as a function of the relative distance $r$. While, at large $r$,
the ground state energy is weakly repulsive and flat, the excited state level is
strongly $r$ dependent and decreases substantially as $r$ becomes small.
By way of extrapolation,
the salient feature of the spectrum is a level crossing of the ground and excited
states at about $r\approx 0.2{\rm fm}$. There, the adiabatic potential changes
from weak repulsion to strong attraction.
Analysis of the eigenvalues of the correlation matrix, at asymptotic times,
reveals that the interaction mechanism gradually changes from being dominated
by quark degrees of freedom (quark-antiquark exchange) at large $r$ to gluon
degrees of freedom( gluon pair exchange) as $r$ becomes smaller. This view is based
on monitoring $\Phi_{1,2}$ operator mixing as a function of $r$.
 
A side aspect of the current work relates to the various types of errors
emerging in the analysis procedure.
Between the gauge configuration statistical errors (which are small), the uncertainties
native to the MEM analysis (which are comparable), and the widths of
the spectral peaks (which are typically large) it remains a matter of judgment to
decide which type of uncertainty is physically relevant.
We have here taken the point of view that the spectral peak width should be
considered the error of the simulation results because it is a measure of
the {\em information content} in the spirit the Shannon-Jaynes entropy contained
in the lattice data. By this measure we accept that the uncertainties generally
exceed the gauge configuration statistical errors.

From the peak volumes of the spectral density functions we learn about the
efficacy of the operators $\Phi_{1,2}$ of coupling to the low-lying physical
excitations of the meson-meson system. Although most of the important
excitation mechanisms appear to be covered by $\Phi_{1,2}$ more operators should
be employed in more detailed studies of hadronic interaction mechanisms.  

Technically, investigations into hadronic interaction are very demanding for both the
need to extract small energy differences and having to dealing with (noisy) nonlocal
operators involving loops that are a combination of gauge link products and light
quark propagators. Anisotropic lattices and advanced analysis techniques
seem essential tools in future studies.

\begin{acknowledgments}
This material is based upon work supported by the National Science Foundation
under Grant No. 0073362.
Resources made available through the Lattice Hadron Physics Collaboration (LHPC)
were used in this project.
\end{acknowledgments}


\end{document}